\documentclass[10pt,aps,prb,superscriptaddress,showpacs,floatfix,nofootinbib,svgnames,eprint]{revtex4-2}

\usepackage{graphicx}
\usepackage{physics}
\usepackage{tikz}
\usetikzlibrary{shapes.geometric}
\usepackage{footnote}
\usepackage[normalem]{ulem}
\usepackage{placeins}
\usepackage{bm}
\usepackage{xcolor}
\usepackage[caption=false]{subfig}
\usepackage{soul}

\newcommand{\be}{\begin{equation}}
\newcommand{\ee}{\end{equation}}
\newcommand{\bn}{\begin{eqnarray}}
\newcommand{\en}{\end{eqnarray}}

\usepackage[colorlinks=true]{hyperref}

\begin{document}

\title{Strange-like Metallicity in a Toy Model with Selective-Mottness}

\author{M. S. Laad$^{1}$ and S. R. Hassan}

\affiliation{ Institute of Mathematical Sciences, Taramani, Chennai 600113, India, and Homi Bhabha National Institute, Anushakti Nagar, Trombay, Mumbai, India}

\date{\rm\today}

\begin{abstract}
  The interplay between band and atomic aspects in materials with co-existing wide-band and flat-band states, or wide-band and effectively dispersionless electronic states is increasingly expected to lead to novel behavior.  Using dynamical mean-field theory (DMFT), we investigate strange-metal-like behavior and emergence of unconventional superconductivity in a toy model that captures this interplay.  Surprisingly, we find good accord with transport features seen in underdoped cuprates and ladder Fe-arsenides.  We connect our findings to proposals of FL$^{*}$ and orthogonal Fermi liquids, and present a route to it's direct instability to novel, competing orders.
\end{abstract}
   

\maketitle

\vspace{1.0cm}

{\bf Introduction}.
 Flat-band electronic systems can manifest as a result of geometric frustration
 or in Moire settings as in the celebrated case of twisted bilayer graphene~\cite{TBLG}.  In general, flat-band states (FBS) occur along with dispersive band states.  While the non-trivial topology associated with FBS has been
the focus of recent studies~\cite{Mohit} in connection with novel superconductive
instabilities, the interplay between electronic correlations and the co-existent dispersive and FB states is of more general interest~\cite{Coleman}.  In many cases, very narrow bands, FB-like in character, hybridize with wider band states: the 
actual microscopic character of this mixing is system-dependent, and can be
rather intricate in momentum space: an example is the situation in $f$-electron systems, where 
a popular model is an Anderson lattice model, with/without an additional
interaction between $f$- and wide-band $d$-fermions.  It is the momentum-dependent one-electron hybridization that mixes the atomic $f$-states and produces a narrow but 
dispersive $f$-like band, and the relevance of this feature for a host of 
novel phenomena in $f$-electron systems is well known for some time (references) as the celebrated Kondo-vs-RKKY competition~\cite{Si}.  Moreover, even in effective
 low-energy descriptions of $d$-band oxides, such ``flat band'' situations can occur in 
{\it effective} models, suitably extracted from more realistic quantum 
chemistry: see, for example, Sire {\it et al.}~\cite{Varma}.

In this paper, we will consider an Anderson-like model in it's strong correlation regime ($U_{ff}=\infty$ as an appropriate fixed point), supplemented by a 
finite $U_{fc}$ and a local or non-local one-electron mixing ($V_{fc}$ or 
$V_{fc}(k)$, see below).  We will examine this model in some detail in various 
regimes: $(i)$ where $V_{fc} (V_{fc}(k)$) is irrelevant, giving the so-called spin-$1/2$ FK model~\cite{Freericks-RMP}, and $(ii)$ where $V_{fc} (V_{fc}(k)$) is relevant at a {\it second order}.  We will show how $(i)$ results in a highly non-conventional metal, reminiscent in many ways of a ``strange'' metal, and discuss how 
$(ii)$ leads to direct instabilities of this metal to novel, competing orders,
inducing either novel superconductivity or novel exciton-condensate driven
 density-wave states.  For $(i)$, we use extant DMFT results and extend them to
incorporate effects missed in earlier work.  While our findings could be 
relevant to cases where FB and dispersive bands co-exist, we will also argue 
that these could be more widely applicable to other cases, including those with topological FBS in an effective 
model sense~\cite{Coleman}.       

\vspace{1.0cm}

{\bf Toy Model and Solution}

\vspace{1.0cm}
 
We consider an extended Anderson (or two-band) model in it's strong correlation limit, and extract an effective model that is exactly 
solvable in high dimensions ($d=\infty$).  It exhibits a {\it local} breakdown 
of Landau's Fermi liquid (LFL) theory.  We find that the metallic state violates Luttinger's theorem, and can be viewed, in many respects, as the real space counterpart of the exactly solvable Hatsugai-Kohmoto (HK) model~\cite{Phillips}.  Unlike the HK 
case, it relies on purely local Hubbard interactions and captures doping- 
or interaction-induced {\it dynamical} spectral weight transfer.  We also find two types of non-Landau quasiparticle excitations.  The first correspond to Hubbard band pseudoparticles that are direct real-space analogues of the composite quasiparticles in the HK model.  The second type correspond to an 
infra-red singular multifermion contiunuum that bears intriguing similarities 
with the contribution of holographic matter in AdS-CFT approaches~\cite{Sachdev2010} to strange metallicity.  Remarkably, this will also turn out to allow 
$(i)$ generation of an infra-red singular 
local one-electron propagator and spin fluctuation spectrum, enabling a 
natural route to a ``strange-like'' metal and $(ii)$ a direct instability to 
a nodal ($d$-wave in $d=2$) preformed pair state.

\vspace{1.0cm}     

      We begin with the extended periodic Anderson model, $H=H_{0}+H_{1}$, where the non-interacting part,

\be
H_{0}=\sum_{k,\sigma}\epsilon_{k,c}c_{k,\sigma}^{\dag}c_{k,\sigma} + E_{f}\sum_{i,\sigma}f_{i,\sigma}^{\dag}f_{i,\sigma} + \sum_{k,\sigma}(V_{fc}(k)c_{k,\sigma}^{\dag}f_{k,\sigma}+h.c)
\ee

contains a non-dispersive $f$-level hybridized with a band of $c$-fermions by 
a (local or non-local) hybridization.  And we have

\be
H_{1}=U_{ff}\sum_{i}n_{i,f,\uparrow}n_{i,f,\downarrow} + U_{fc}\sum_{i}n_{i,f}n_{i,c}
\ee
with $n_{i,b}=\sum_{\sigma}n_{i,b,\sigma}$ and $b=f,c$.  

In reality, the $c,f$ orbitals can represent $p,d$ or $d_{x^{2}-y^{2}},d_{z^{2}}$ in transition-metal oxides, $p(d),f$ states in rare-earth compounds, or nodal ($N$) and anti-nodal ($AN$) states in cuprates.  The detailed form of $V_{fc}(k)$ is slaved to the local quantum chemistry of the system in each case. 

Taking the $U_{ff}=\infty$ limit excludes the upper Hubbard band (UHB) in the
$f$-sector via a projector which forbids double $f$-occupancy at a site.  Thus,
we must replave $f_{i,\sigma}\rightarrow (1-n_{i,f,-\sigma})f_{i,\sigma}$ in 
$H$ above.  This complicates the analysis, so we proceed as follows.

We appeal to the concept of ``hidden fermions'' introduced by Zhu {\it et al.}~\cite{zhu} and by Imada {\it et al.}~\cite{Imada}.
We write $f_{i,\sigma}=F_{i,1,\sigma}+F_{i,2,\sigma}=(1-2n_{i,f,-\sigma})f_{i,\sigma} + 2n_{i,f,-\sigma}f_{i,\sigma}$, and notice that the second component is projected out at $U_{ff}=\infty$.  In the strict no $f$-double occupancy limit, the first component can be 
written in a variety of ways, as $F_{i,1,\sigma}=(-1)^{n_{i,f,-\sigma}}f_{i,\sigma}=s_{i}^{z}f_{i,\sigma}$, with $s_{i}^{z}$ a fluctuating $Z_{2}$ 
Ising degree of freedom.  Importantly, $F_{1,\sigma}$ is also orthogonal to $f_{\sigma}$ in this limit.  In contrast to the Hubbard operators, we see that the $F_{i,1,\sigma}$ satisfy the usual fermionic anticommutation relations in the no-double $f$-occupancy subspace.  Now, $H$ takes the form

\be
H_{0}=\sum_{k,\sigma}\epsilon_{k,c}c_{k,\sigma}^{\dag}c_{k,\sigma} + E_{f}\sum_{i,\sigma}F_{i,1,\sigma}^{\dag}F_{i,1,\sigma} + \sum_{k,\sigma}(V_{fc}(k)c_{k,\sigma}^{\dag}F_{k,1,\sigma}+h.c)
\ee

where $F_{k,1,\sigma}=\sum_{q}f_{k+q,\sigma}(\delta_{q,0}-2n_{q,-\sigma})$, and

\be
H_{1}=U_{fc}\sum_{i,\sigma,\sigma'}F_{i,1,\sigma}^{\dag}F_{i,1,\sigma}n_{i,c,\sigma'}
\ee

It is well known from extensive studies that in symmetry unbroken phases (no 
magnetic, charge, superconductive order), the metallic ground state is either
$(1)$ a severely renormalized Landau Fermi liquid when the hybridization is RG
relevant, since the $f$-local moment is eventually ``Kondo'' screened by the
$c$-Fermi sea spin density as an eventual consequence of a second-order-in hybridization process, or $(2)$ the $f$-moments form dynamically fluctuating short-range valence bond spin singlets, via an effective exchange induced by the
second-order-in $V_{fc}$ process upon {\it irrelevance} of $V,V(k)$ at one-electron level, whence no Kondo-induced
Landau FL can obtain.  In fact, this is the fractionalized Fermi liquid (FFL) state~\cite{Sachdev-FFL}.  In the symmetry-unbroken phase, this leads to either a local moment metal, or to a dimerized spin liquid co-existing with itinerant carriers, depending upon extent of geometric frustration.

We begin by considering the consequences arising from a regime where the
hybridization, $V_{fc}(k)$, is RG irrelevant at one-fermion level.  This is the regime where a finite $V_{fc}$ cannot coherently mix $c$ and $F_{1}$ fermions in the OSMP: the latter generically occurs in multi-orbital Hubbard~\cite{vonDelft,Laad2016} or extended-PAM~\cite{Laad2012}.  We discuss the resulting
fluctuating local moment metal in detail, and describe it's resistive response.  Surprisingly, we find very good qualitative accord with data for TBLG, underdoped cuprates and pressurized BaFe$_{2}$S$_{3}$.  We also show how such a ``strange''-like metal can undergo direct instabilities to a range of competing ordered states ($d$-wave in $D=2$), depending upon local quantum chemical details.

When the hybridization is RG irrelevant at one-fermion level, we can discard
it in $H$ above.  The resulting model is the spin $S=1/2$ Falicov-Kimball model, where the $F_{1,\sigma}$ fermions are {\it effectively} immobile. 

\be
H_{FK}=-t\sum_{<i,\delta=e_{x},e_{y}>}(c^{\dag}_{i\sigma}c_{i+\delta,\sigma}+h.c) + U_{fc}\sum_{i,\sigma,\sigma'}n_{i,c,\sigma}n_{i,F_{1},\sigma'} -\mu\sum_{i,\sigma}(n_{i,c,\sigma}+n_{i,F_{1},\sigma})
\ee

      A feature of the toy model is that $(i)$ it is a model with FBS and dispersive band states that strongly interact with each other (by construction, a ``two-fluid'' model), and $(ii)$ the $F_{1}$-fermions present a local, disordered (in the paramagnetic state) scattering potential 
for the mobile $c$-fermions.  But the localized $F_{1}$-fermions experience a 
time-dependent, ``suddenly switched on (off)'' local
potential due to the itnerant $c$-fermions, on a time-scale 
$\tau\simeq \hbar/t_{c}$, in the manner of a sudden, local quantum quench.  
  
\vspace{1.0cm}

{\bf Exact DMFT Solution of the spin-$1/2$ Falicov-Kimball Model: The ``Alloy Analogy''}

\vspace{0.5cm}
  
     The exact local Green function for our simplified Hubbard (or FK) model 
can be readily written down by a direct and repeated application of the equation-of-motion technique~\cite{Freericks-RMP}.                                              

\be
G_{ii,c}=\frac{1-<n_{i,F_{1}}>}{\omega -t^{2}G_{ii,c}} +
\frac{<n_{i,F_{1}}>}{\omega - U_{fc} -t^{2}G_{ii,c}}
\ee
 
      For a Bethe lattice, the Mott 
transition occurs at $U_{fc}\simeq O(W)$, the non-interacting one-electron 
band-width.
 The local density-of-states (DOS) is a superposition of lower-Hubbard band (LHB) and upper Hubbard band (UHB) states, represented by the Hubbard operators
 $X^{0\sigma}_{i}=(1-n_{i,F_{1}})c_{i,\sigma}$ and $X^{-\sigma\sigma}_{i}=n_{i,F_{1}}c_{i,\sigma}$ respectively.  It is noteworthy that the ``hidden fermion'' 
that is of interest 
 in certain theories for the $d$-wave PG for cuprates is simply the difference 
of the above lower-  and upper Hubbard band operators: $C_{i\sigma}=(X^{0\sigma}_{i}-X^{-\sigma\sigma}_{i})=(1-2n_{i,F{1}})c_{i,\sigma}$, and thus has an 
 unbreakable link to Mottness.  Given that the spectral function above is a
 superposition of lower- and upper Hubbard band states, the spectral function 
of the dark fermions must also be finite at the Fermi energy in the metal.

      Thus, the original fermions are ``{\it fractionalized}'' 
into lower- and upper Hubbard band states.  These are the true excitations, 
and the original fermions constitute the ``hidden Fermi liquid''.  Now, there
 is no remnant of any single electron/hole-like Landau quasiparticle states at 
the Fermi surface, simply because the low-energy states now correspond to a 
composite of an electron and a hard-core bosonic local spin fluctuation 
(this is seen most easily by rewriting, e.g, 
$X^{0\sigma}_{i\sigma}=F_{i,1,-\sigma}F^{\dag}_{i,1,-\sigma}c_{i,\sigma}$,
 and similarly for $X^{-\sigma\sigma}_{i}$).  Interestingly, these are 
exactly the real-space analogues of the composite excitations in the Hatsugai-Kohmoto model~\cite{Phillips}.  The corresponding local self-energy is

\be
\Sigma_{c}(\omega)=U_{fc}<n_{i,F_{1}}> + \frac{U_{fc}^{2}<n_{i,F_{1}}>(1-<n_{i,F_{1}}>)}{\omega +\mu -U_{fc}(1-<n_{i,F_{1}}>)-t^{2}G_{c}(\omega)}
\ee
Beyond a critical $U_{fc}=U_{c}>W$, one obtains a Mott insulator 
characterized by a {\it zero} of the one-electron propagator.  Thus, the 
Fermi surface of the non-interacting model (surface of poles of 
$G_{c,k}(\omega)$) is supplanted by a Luttinger surface (surface of zeros of 
$G_{c,k}(\omega)$) in the insulator, and Im$\Sigma_{c}(\omega)$ develops at pole at the Fermi surface.  In the metal, Im$\Sigma_{c}(\omega=E_{F})$ is 
always finite, and this clearly shows that the symmetry-unbroken metallic 
phase is never a LFL for {\it any} $U_{fc}/t$ off half-filling.  Thus, 
Luttinger's theorem is always violated, and low-energy single electron/hole 
like excitations are always unstable, because they decay into composite 
Hubbard band excitations before they can be registered as fundamental 
long-lived quasiparticles.  Because the low-energy states are composites 
of single (unprojected) electrons plus local $F_{1}$-fluctuations, the 
total spectral weight at low energy must be larger than that from a naive counting, since the kinetic energy operator connects the above lower- and upper Hubbard band states.  This results
 in the additional {\it dynamical} weight of $O(t/U_{fc})$ coming from these 
local fluctuations.  

\vspace{1.0cm}

{\bf Infra-Red ``Strange-Metal-like'' Singularities}

\vspace{0.5cm}

   The usual DMFT solution of the FKM also shows up a crucial aspect: the local 
 $F_{1}$-fermion dynamics is highly non-trivial.  In presence of a $c$-fermion 
Fermi sea, the effect of a $U_{fc}$ translates into a ``sudden switching-on'' 
(or, in modern parlance, a sudden local quench) of a localized potential due to
 a $c$-electron hopping on and off on a time scale $\hbar/t$ on any given site 
$i$, as seen by a localized $F_{1}$-electron as a function of time.  But this
 is just the lattice version of the venerated X-ray edge problem, and has dramatic consequences, as Anderson has repeatedly emphasized~\cite{Anderson}.  In DMFT, this is 
{\it not} true for the usual Hubbard model, but it indeed rigorously holds for 
the (spinless or spinful) FKM.  Specifically, this process implies generation of an infinite number 
of local particle-hole (spin-excitonic in the FK model case) excitations in 
response to a ``sudden'' (local) quench, induced by $U_{fc}$.  Thus, the 
correlator of these local ``excitons'', made up from a $c$-electron and $F_{1}$-hole, and written as $(c_{k,\sigma}F_{1,k,\sigma}+h.c)$ in momentum space, 
turns out to be infra-red singular with a fractional, interaction-dependent 
exponent!  Moreover, the $F_{1}$-fermion spectrum also picks up this same 
singularity.  We have~\cite{GKS},

\be
Im G_{F_{1}}(\omega)\simeq \frac{(1-n_{F_{1}})\theta(\omega+\mu)+ n_{F_{1}}\theta(-\omega)-\mu}{|\omega+\mu|^{1-\eta}}    
\ee

and,

\be
Im \chi_{cF_{1}}^{+-}(\omega)= Im \int dt e^{i\omega t}<T[c_{i,\sigma}^{\dag}F_{1,i,\sigma}(t);F_{1,i,\sigma'}^{\dag}c_{i,\sigma'}]> \simeq \frac{(1-n_{F_{1}})\theta(\omega+\mu) + n_{F_{1}}\theta(-\omega-\mu)}{|\omega+\mu|^{2\eta_{\sigma\sigma'}-\eta_{\sigma\sigma'}^{2}}}
\ee
where $\eta_{\sigma\sigma'}=\frac{1}{\pi}$tan$^{-1}(U_{fc}^{\sigma\sigma'}/W_{c})\simeq (U_{fc}^{\sigma\sigma'}\rho_{c}(E_{F}=0))$ (the second limit applies only at small $U_{fc}/W$)
 with $W_{c}=2zt$ is the non-interacting $c$-electron bandwidth.  In the above eqn, we have allowed for the possibility that the interband interaction in 
 real systems obeys $U_{fc}^{-\sigma\sigma}\neq U_{fc}^{\sigma\sigma}$ because of a finite Hund coupling.  In the Supplementary Information (SI), we present a derivation leading to an indication 
 of this singular behavior in $G_{F_{1}}$ for the FKM using equations-of-motion technique.  

     Using the relation
$G_{F_{1}}^{-1}(\omega)=\omega -\Sigma_{F_{1}}(\omega)\simeq \omega^{1-\eta}$, 
the corresponding $F_{1}$-fermion self-energy is just $\Sigma_{F_{1}}(\omega)\simeq -\omega^{1-\eta}$ at low energy.  It is interesting to 
notice that an exact DMFT computation of $G_{F_{1}}(\omega)$ using the
numerical renormalization group (NRG) solver~\cite{Anders} is fully consistent with the above form, up to high energy $O(W)$.  Since we need $G_{F_{1}},\Sigma_{F_{1}}$
to have correct large-$\omega$ behavior, we modify the above form for $\Sigma_{F_{1}}(\omega)$ by the replacement

\be
Im\Sigma_{F_{1}}(\omega)=-U_{fc}^{2}n_{c}(1-n_{c})\frac{|\omega+\mu|^{1-\eta}\Omega_{c}^{\eta}}{(\omega+\mu)^{2}+W^{2}}
\ee

 by hand: Here, we added $\Omega_{c}^{\eta}$ as a cut-off to restore the correct dimension for the $F_{1}$-fermion self-energy.  Of course, this is an approximation, but in good accord at
 low-to-intermediate as well as large energy when compared to DMFT (NRG) results.  Analytically, similar results 
(but there restricted to weak coupling and close to the Fermi surface) obtain 
from bosonization~\cite{Schotte} as well as from parquet functional RG~\cite{Sharma}
 analyses of the underlying impurity problem.  Moreover, in a phenomenological 
 vein, Leong {\it et al.}~\cite{Phillips-unparticles} use the power-law self energy to find a density-of-states having a 
power-law singular form, $\rho(\omega)\simeq |\omega|^{-\alpha}$, for 
$\alpha < 1/2$ (or for $0<\eta <1$ for the FKM).  The difference between our
work and that of Leong {\it et al.} is that unparticles are selectively Mott
 localized, effectively dispersionless ``composite fermions'' in our case.

      At finite $T$, one of course needs to replace the $T=0$ form of 
$G_{i,i,F_{1}}(\omega)=G_{F_{1}}(\omega)$ by 

\be
G_{F_{1}}(\omega,T)=e^{i(\phi+\pi(1-\eta)/2)}T^{-(1- \eta)}\frac{\Gamma(\frac{\eta}{2}-\frac{i\omega}{2\pi T})}{\Gamma(1+\frac{\eta}{2}-\frac{i\omega}{2\pi T})}
\ee

 with $0<\eta<1$.  This is
the same form as the contribution of the ``holographic sector'' in AdS/CFT
approaches to strange metals~\cite{Sachdev2010}.  In our case, it is the response of the ``selectively'' localized composite fermion $F_{i,1,\sigma}=(1-2n_{i,f,-\sigma})f_{i,\sigma}$, and arises from lattice X-ray edge physics in DMFT.

\vspace{1.0cm}

{\bf Solution Including ``Beyond Alloy Analogy'' Effects}

\vspace{0.5cm}

Close examination of the structure of the above solution presents a difficulty.
The structure of $G_{c}(\omega)$ assumes that the $n_{i,F_{1},\sigma}$ act 
as static, (strong) potential scatterers for the $c$-fermions (in fact, this is the famous ``scattering correction'' of Hubbard~\cite{Hubbard1964}, equivalent to 
the best single-site-theory, the CPA, for disordered binary alloys).  But as we
have seen, the $F_{1}$-fermion spectral function shows a non-trivial branch-point structure in the infra-red.  Hence, the assumption of the $F_{1,\sigma}$ fermions 
presenting a static, random, alloy potential for the $c$-fermions cannot be 
correct.  The singular fluctuations of the $F_{1,\sigma}$ can be regarded as 
singular ``valence'' fluctuations, and must drastically modify both, the
$c$ and $F_{1}$-fermion responses in a self-consistent way.  To our best knowledge, such ``beyond AAA'' effects have never been considered in earlier work on the (spinless or spinful) FK model, where the dynamical feedback of the $F_{1}$-sector on the $c$-sector is absent.

We thus realize that we must now allow the dynamical feedback of the 
$F_{1}$-fermion spectral function into $G_{c}(\omega)$ and vice-versa in a 
fully self-consistent way.  Since $G_{c}(\omega), G_{F_{1}}(\omega)$ are both
non-quasiparticle-like, it follows that the $c$ and $F_{1}$-fermions cannot ``see each other'' as one-fermion-like quasiparticles in the intermediate state during any scattering 
process.  This allows us to neglect the irreducible vertex corrections, and to employ a dynamical $1/N$-like approximation to 
treat the feedback effects mentioned above in a selfconsistent way that corrects the above difficulty.

      In the dynamical $1/N$ approach, we use the above forms of $G_{c},G_{F_{1}}$ (or $\Sigma_{c},\Sigma_{F_{1}}$) as an initial input choice.  The leading 
order-in-$1/N$ contributions to the corresponding self-energies then read

\be
\Sigma_{c}^{(N)}(\omega)=U_{fc}^{2}\int\frac{d\epsilon}{\pi}Im\chi^{F_{1}F_{1}}(\epsilon)G_{c}(\omega-\epsilon)\frac{1+f(\epsilon)-f(\omega-\epsilon)}{\omega-\epsilon}
\ee

with $\chi^{F_{1}F_{1}}(\omega)=\int\frac{d\epsilon}{\pi}G_{F_{1}}(\omega+\epsilon)G_{F_{1}}(\epsilon)$.  Similar equations with $c,(F_{1})$ replaced by $F_{1},(c)$ hold for the $F_{1}$-electron self-energy.  We notice that this ansatz is very similar to the ``non-crossing approximation'' (NCA)~\cite{Mueller-Hartmann} that has been used quite successfully as an impurity solver for DMFT as long as no Landau Fermi liquidity is 
expected to occur.  For our $S=1/2$ FKM, this is indeed true, as both, usual
DMFT~\cite{Freericks-RMP} and our results (see below) will show.  To implement this scheme, we now use the usual DMFT expression for $G_{c}(\omega)=G_{c}^{(0)}(\omega)$, along with our IR-singular ansatz for $G_{F_{1}}(\omega)=G_{F_{1}}^{(0)}(\omega)$ in 
the earlier section as initial guesses.  Then, with the large-$N$ corrections,
 the full Green functions are computed from Dyson's equation.  We have

\be
G_{c}^{-1}(\omega)=[G_{c}^{(0)}]^{-1}(\omega)-\Sigma_{c}^{(N)}(\omega)
\ee

and

\be
G_{F_{1}}^{-1}(\omega)=[G_{F_{1}}^{(0)}]^{-1}(\omega)-\Sigma_{F_{1}}^{(N)}(\omega)
\ee

These expressions form the substance of our self-consistency scheme.  Beginning with the above guesses, we compute the large-$N$ self-energies, get updated guesses for $G_{c},G_{F_{1}}$ from Dyson's equation, and iterate the scheme till 
numerical self-consistency obtains.  Using the converged Green functions, we 
evaluate the $dc$ conductivities from the usual DMFT formulation for both $c,F_{1}$
channels.  Given these two channels, the conductivities add up, and the $dc$ resistivity is

\be
\rho(T)=[\sigma_{c}(0,T)+\sigma_{F_{1}}(0,T)]^{-1}
\ee

\vspace{1.0cm}

{\bf Results}

\vspace{1.0cm}

\begin{figure}[htbp]
\centering
\begin{minipage}{0.49\textwidth}
    \centering
    \includegraphics[width=\textwidth]{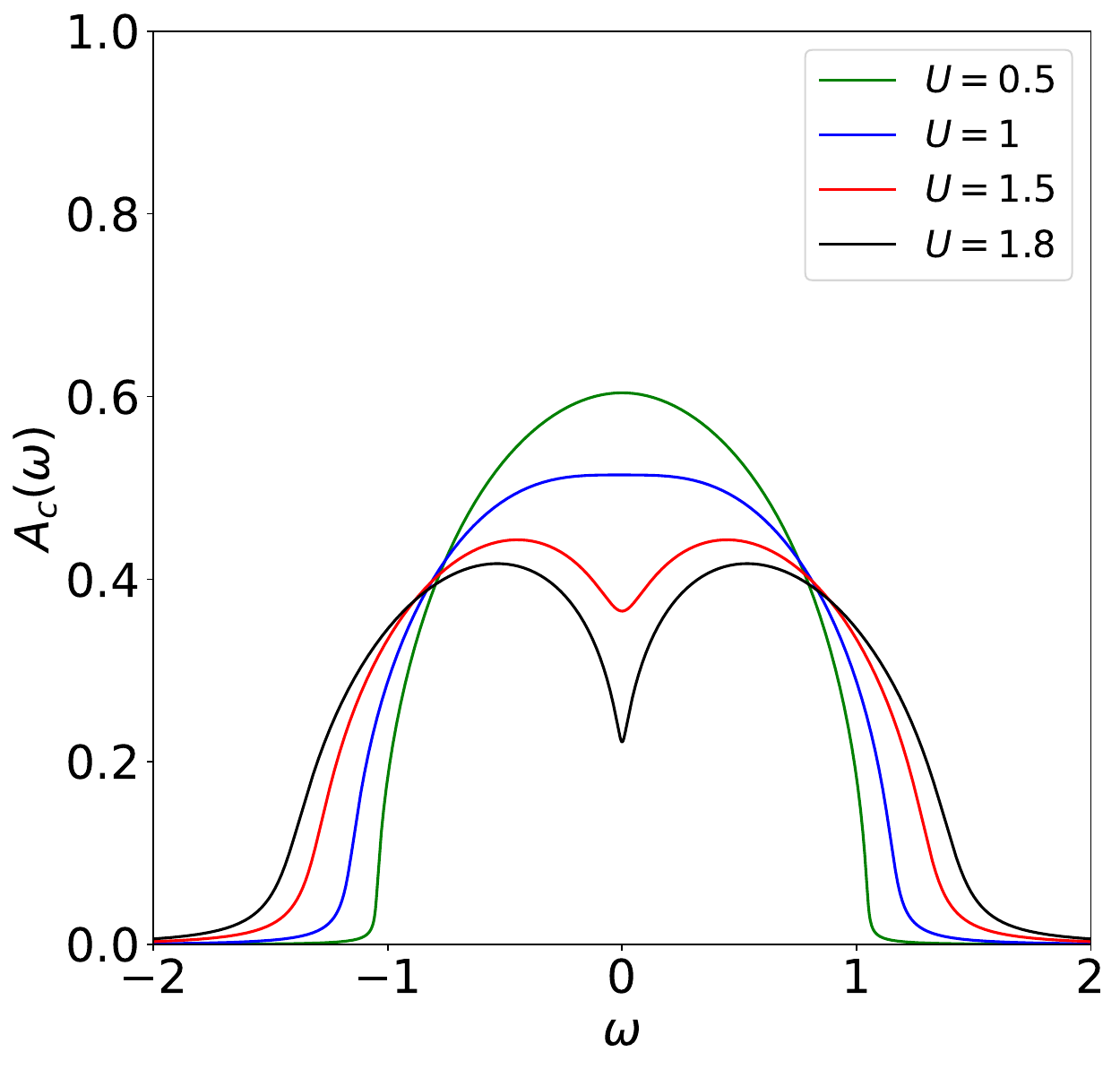}
\end{minipage}
\hfill
\begin{minipage}{0.49\textwidth}
    \centering
    \includegraphics[width=\textwidth]{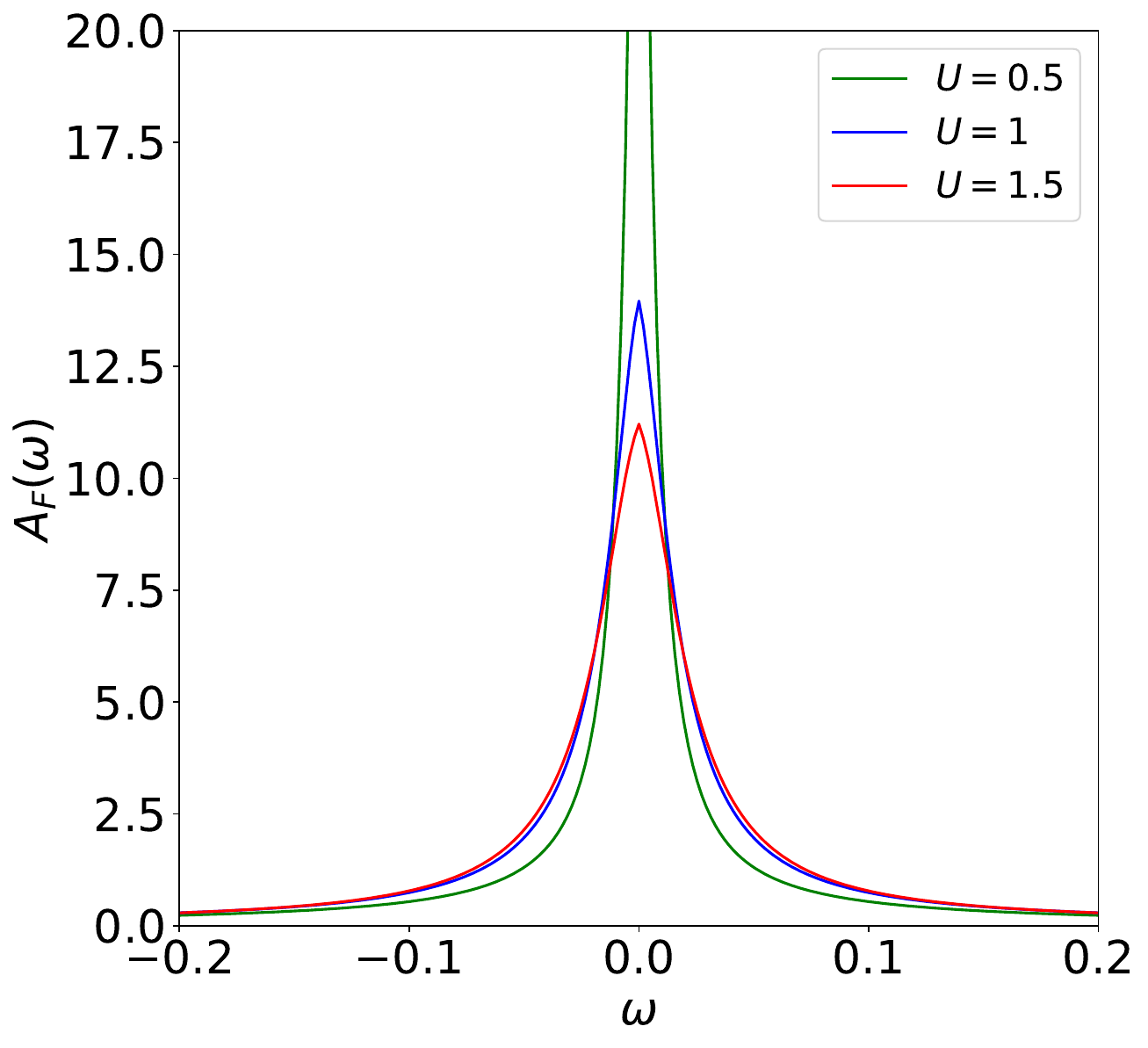}
\end{minipage}
\vspace{0.2cm}
\caption{Local spectral functions (DOS) for $c$ fermions (left) and $F_{1}$ fermions (right) across various interaction strengths $U$, in the symmetry-unbroken metallic phase at fixed temperature $T = 0.05$. As $U$ increases, $\rho_{c}(\omega)$ exhibits a deepening pseudogap, while $\rho_{F_{1}}(\omega)$ develops a broadened lattice X-ray edge singularity}
\label{fig1:cfDOS}
\end{figure}

We now describe our results.  In Fig.~\ref{fig1:cfDOS}, we show the $c,F_{1}$-fermion 
local spectral functions.  In the alloy-analogy approximation (AAA), 
$G_{c}$ is $T$-independent, so that any $T$-dependence in transport, 
etc, arises from the Fermi-Dirac distribution function or its energy derivative.  In stark contrast, we find that going beyond the AAA introduces a non-trivial
$T$-dependence: it is small but noticeable in $\rho_{c}(\omega,T)$ but much 
more pronounced in $\rho_{F_{1}}(\omega,T)$.  In particular, the IR-singularity in
$\rho_{F_{1}}(\omega,T=0)$ is appreciably broadened with increasing $T$ (not shown).  See, however, Fig.~\ref{fig4:scaling_f}).  Heuristically, we can understand this as a thermally induced broadening of the lattice 
version of the Nozieres-de Dominicis X-ray edge singularity.  The term $U_{fc}$
``fills'' the ``core hole'' $F_{1}$-fermion state with a higher probability as $T$ increases due to thermal enhancement of electron-hole excitations of the $c$-fermion
``Fermi sea''.  The effect of this is to introduce a $T$-dependent lifetime factor, $\Gamma(T)$ that smears the singularity in $\rho_{F_{1}}(\omega)$ {\it a la} Doniach-Sunjic~\cite{Doniach-Sunjic}.

\begin{figure}[htbp]
\centering
\begin{minipage}{0.49\textwidth}
    \centering
    \includegraphics[width=\textwidth]{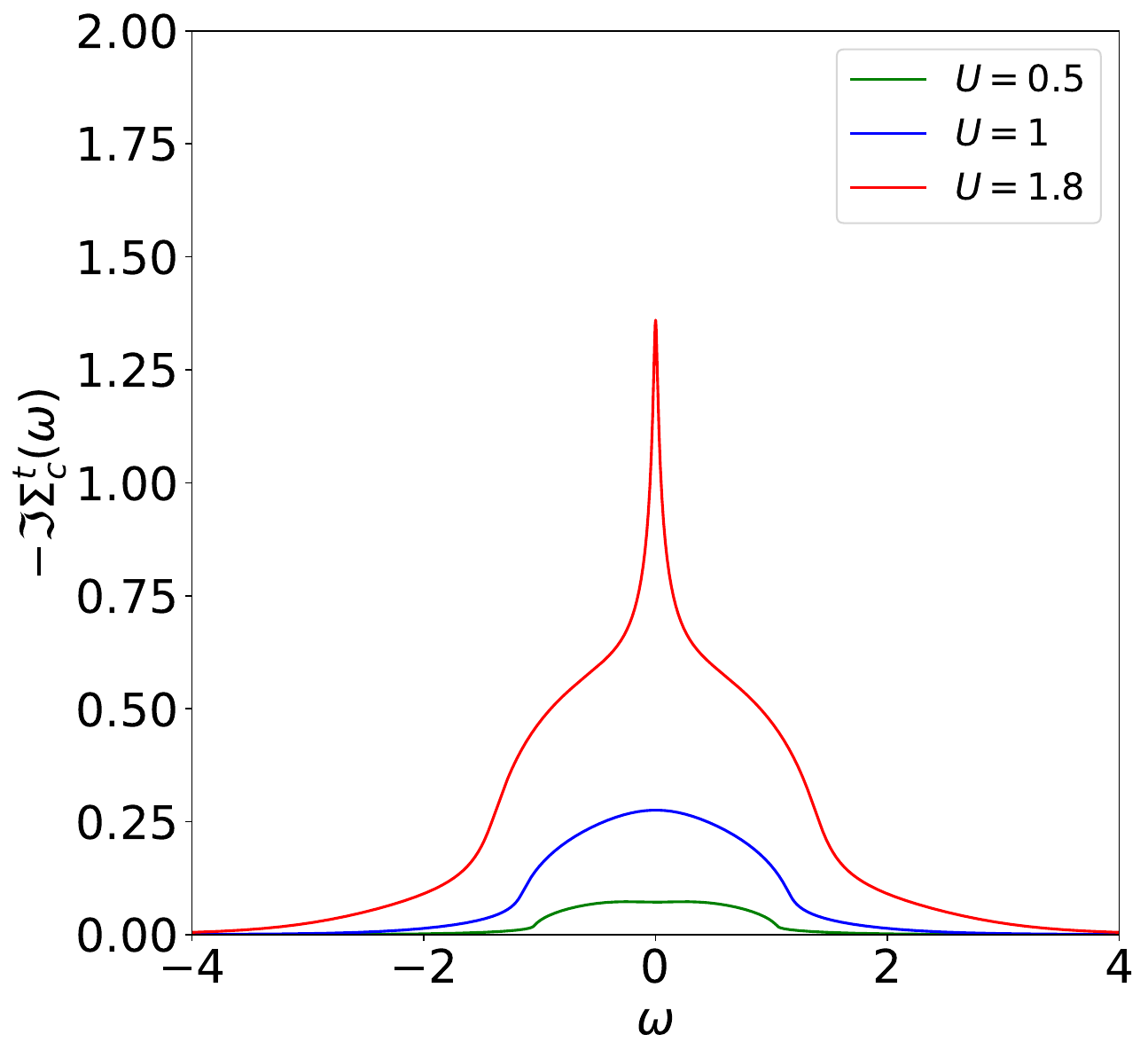}
\end{minipage}
\hfill
\begin{minipage}{0.49\textwidth}
    \centering
    \includegraphics[width=\textwidth]{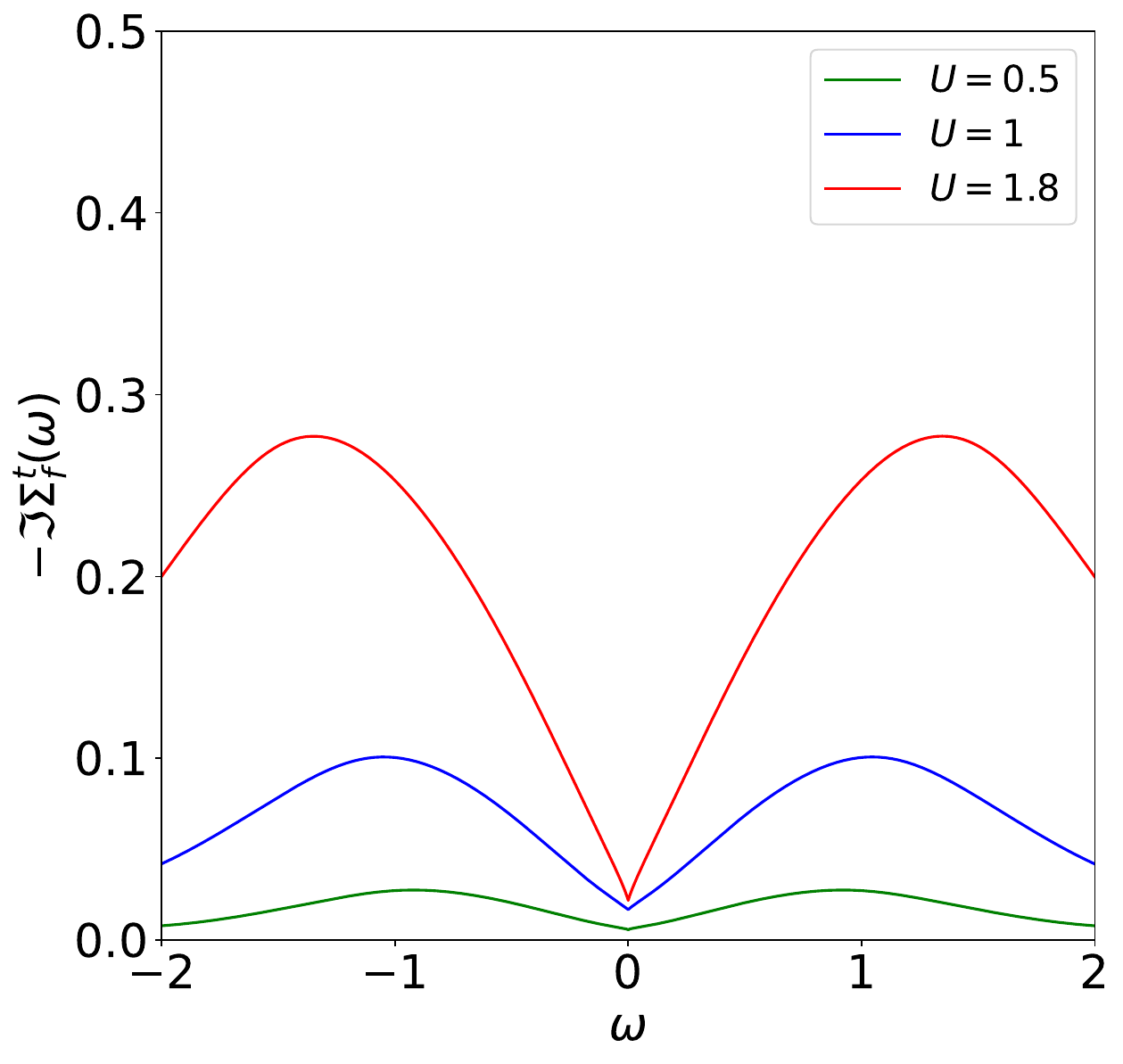}
\end{minipage}
\vspace{0.2cm}
\caption{The total self energies  for $c$ fermions (left) and $F_{1}$ fermions (right) across various interaction strengths $U$, in the symmetry-unbroken metallic phase at fixed temperature $T = 0.05$.}
\label{fig2:cfSELF}
\end{figure}

In Fig.~\ref{fig2:cfSELF}, we exhibit the $c,F_{1}$-fermion self-energies.  While 
Im$\Sigma_{c}(\omega)$ qualitatively retains it's usual FKM structure and
implies a $c$-fermion pseudogap, Im$\Sigma_{F_{1}}(\omega)$ acquires a nearly
linear-in-$\omega$ form at low-energy with an $\omega=0$ kink.  The corresponding real part of the $F_{1}$-fermion self-energy is Re$\Sigma_{F_{1}}(\omega)\simeq \omega$ln$(\omega/\omega_{c})$ with $\omega_{c}$ an appropriate cut-off.
Thus, the metallic state is still a non-Landau FL metal.  The corresponding Landau quasiparticle
residue, $z_{F_{1}}(\omega)\simeq -(ln\omega)^{-1}$, vanishes at the Fermi surface.  Now, the emergent picture 
is interesting: the metal is a two-component or ``two-fluid'' type, with
incoherent $c$-fermions co-existing with a strange metal 
(marginal-FL)~\cite{Varma} like component.  The latter arises from the unquenched, local spin and charge fluctuations associated with a selectively (Mott) localized composite fermion.
Thus, this metal is reminiscent of the FFL liquid.  In fact, because the $F_{1}$-fermion is orthogonal to $f$, this is also an {\it orthogonal Fermi liquid}~\cite{Senthil(OFL)}.  Given that the IR-singular component might be expected to dominate, one might naively expect this channel to dominate and linear-in-$T$ resistivity to result.  However, the actual 
outcome is much more interesting, as we now show.

\vspace{1.0cm}

First, we point out that the marginal-like $F_{1}$-fermion self-energy directly implies a 
$T$-dependent effective mass enhancement that is reflected in the low-$T$ specific heat.  We 
find that $C_{el}(T)=\gamma T=(m^{*}/m)T=-T$ln$T$, yielding $\gamma(T)=-$ln$T$: this is indeed seen 
in the strange metal in cuprates~\cite{logT}.  But there is an additional component arising from the 
$c$-fermions.  This is also of a non-Landau quasiparticle origin, since Im$\Sigma_{c}(\omega)$ has 
the ``wrong'' sign (a minimum instead of a maximum) at low energy.  Emergence of the PG in $\rho_{c}(\omega)$ will, however, cut off the -ln$T$ contribution to the low-$T$ specific heat at low $T$.

\begin{figure}[htbp]
\centering
\begin{minipage}{0.49\textwidth}
    \centering
    \includegraphics[width=\textwidth]{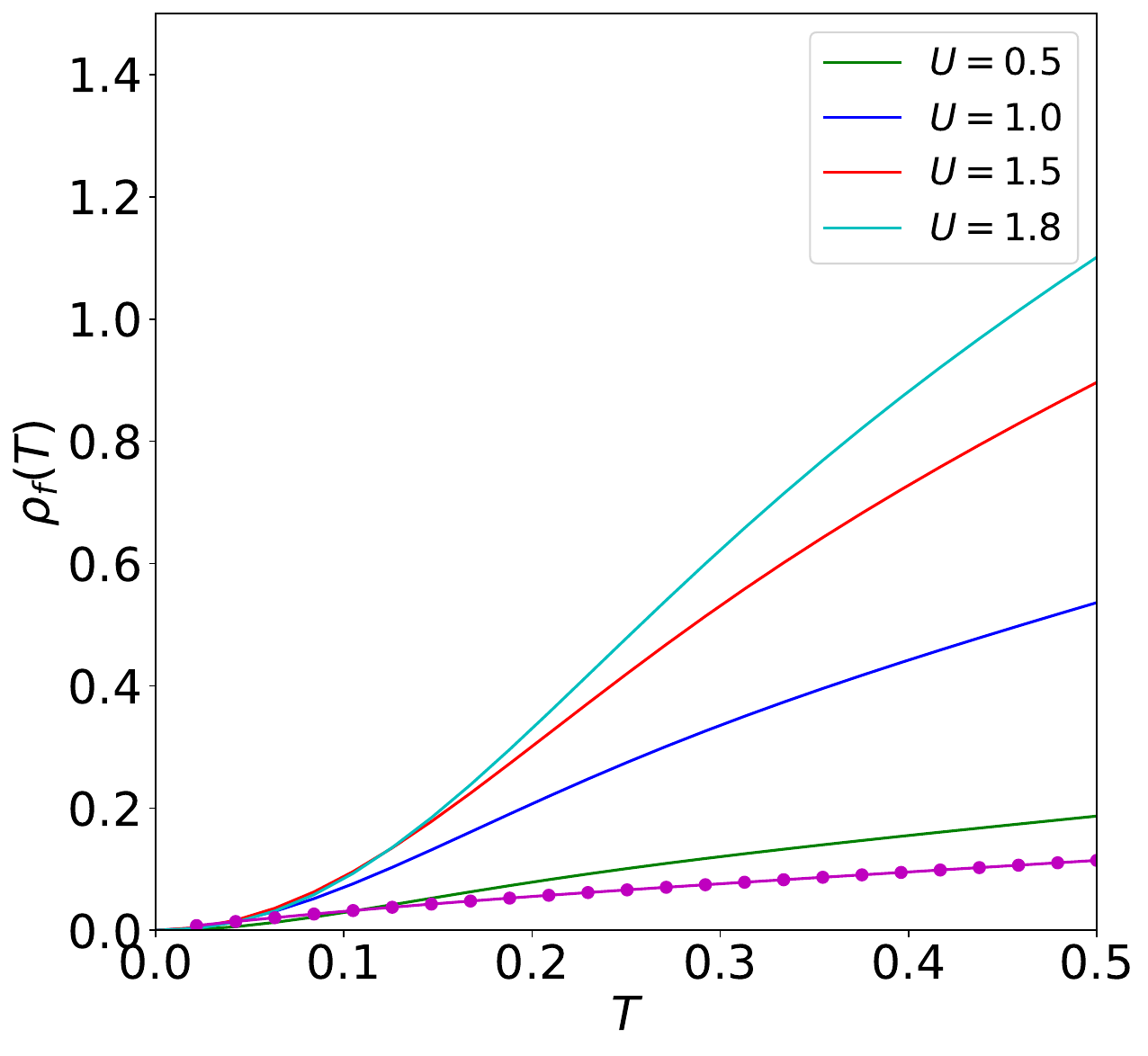}
\end{minipage}
\hfill
\begin{minipage}{0.49\textwidth}
    \centering
    \includegraphics[width=\textwidth]{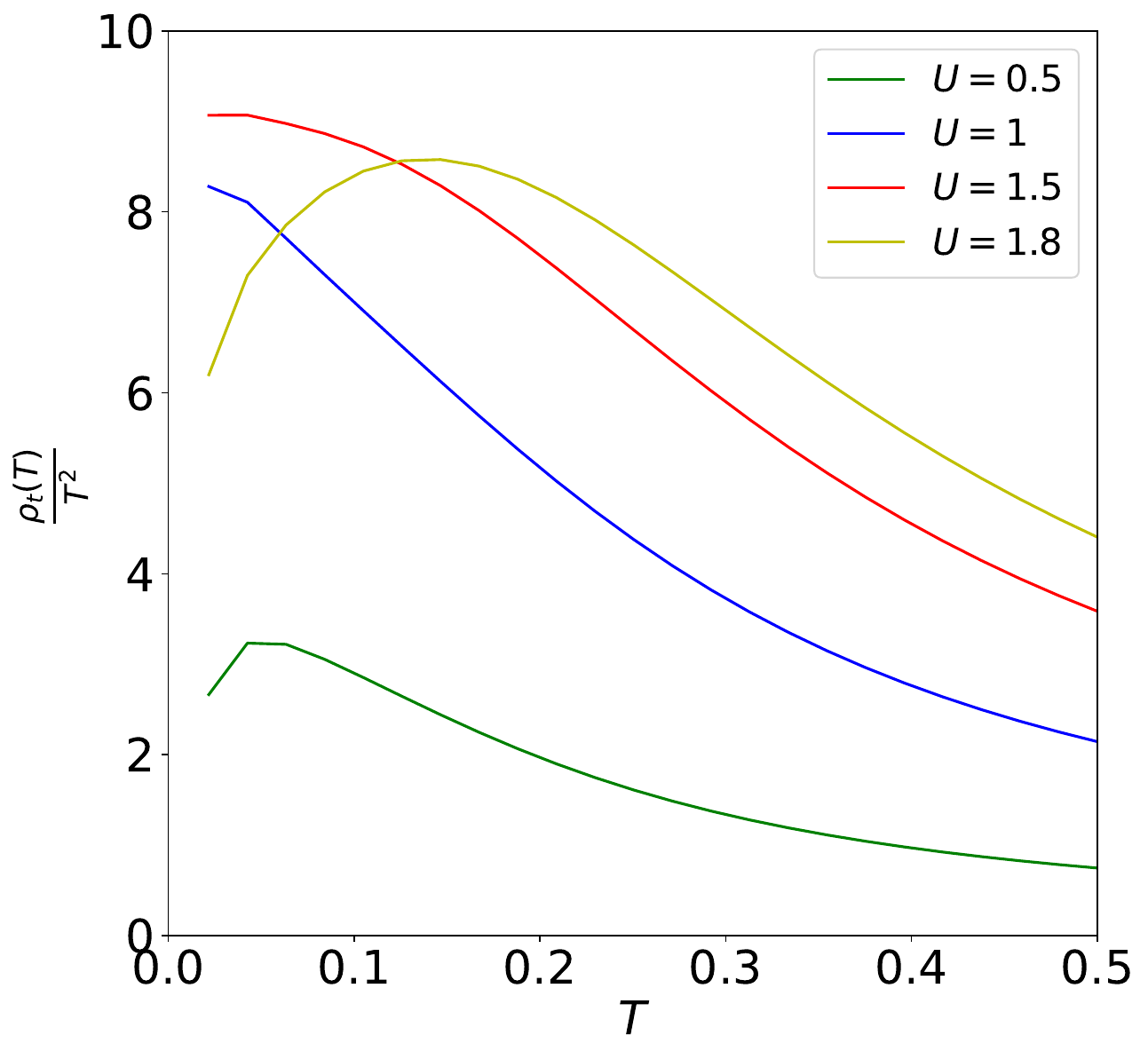}
\end{minipage}
\vspace{0.2cm}
\caption{The left panel shows $\rho_t(T)$ versus $T$ for various values of $U$; the dotted line corresponds to $U = 0.5$ without the dynamical-$1/N$ corrections to the self-energies. The right panel displays $\rho_t/T^2$ plotted against $T^2$}
\label{fig3:rest}
\end{figure}

We now exhibit the total $dc$ resistivity of this two-fluid metal in Fig.~\ref{fig3:rest}(left).  For all $U_{fc}/W$, $\rho(T)$
at high $T$ is bad-metallic and follows a linear-in-$T$ law, extrapolating to a very low value as $T\rightarrow 0$.  As $T$ is reduced, however, a smooth 
crossover with a wide ($U_{fc}/W$-dependent) crossover region occurs, between 
this ``strange'' metal, via a second linear-in-$T$ regime, to a much more ``Fermi liquid-like'' regime at low 
(again $U_{fc}/W$-dependent) $T\simeq (0.01-0.03)$.  This is surprising at first sight, since 
the spectral functions are clearly that of a non-FL metal.  Replotting this
 with $\rho(T)/T^{2}$ in Fig.~\ref{fig3:rest}(right) versus $T$  reveals that eventual $\rho(T)\simeq T^{n}$
behavior, with $n\geq 2$ obtains at lower $T$, For $U_{fc}/W=1.8$, close to the Mott transition, a clear maximum at $T\simeq 0.137$, and a smooth crossover to a $T^{n}$-like form at very low $T\leq 0.01-0.05$ for smaller $U_{fc}$ hint at the
 ``hidden'' influence of the $c$-fermion pseudogap.  This is because this $T$
scale reduces with reduction in $U_{fc}$: linear-in-$T$ resistivity obtains 
over a progressively wider $T$ range, up to lower $T$ as $U_{fc}$ (and hence 
the $c$-fermion pseudogap) is reduced.  However, this state 
is not a Landau FL metal, as discussed before.  We are thus better off interpreting this behavior in terms of a metal where the $c$-fermion PG cuts off the IR singularity in the $F_{1}$-sector, reinstating low-$T$ quasicoherence.

\begin{figure}
    \centering
    \includegraphics[width=0.5\linewidth]{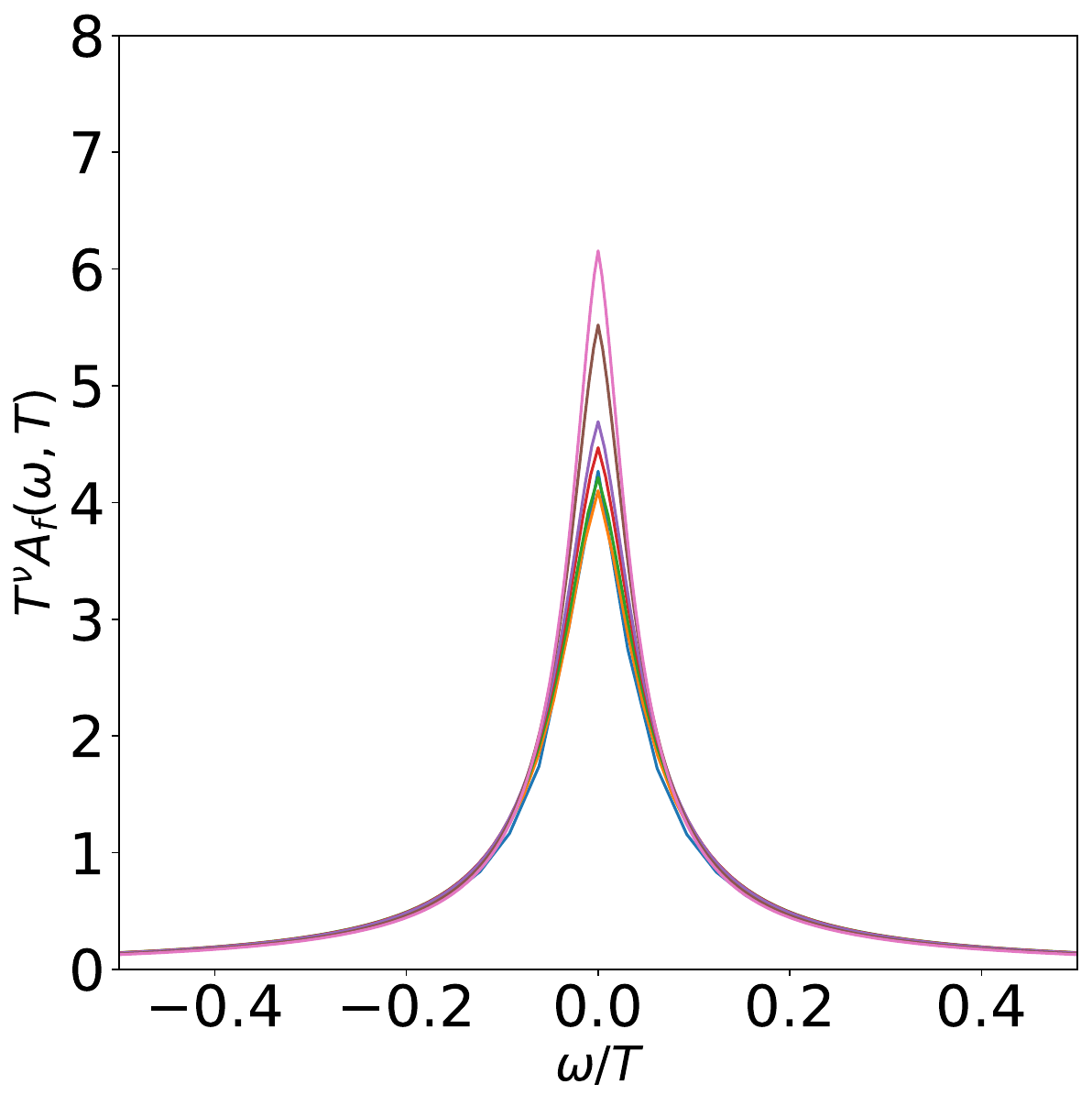}
    \caption{$T^{\nu} A_{F_{1}}(\omega)$ with $\nu=(1-\eta)$ plotted as a function of $\frac{\omega}{T}$ for $U = 1.0$, shown at temperatures $T = 0.06,\ 0.12,\ 0.18,\ 0.24,\ 0.4,\ 0.5$.  At high $T$, $T^{\nu}A_{F_{1}}(\omega,T)=G(\omega/T)$ is perfectly 
    obeyed.  At lower $T$, $c$-fermion pseudogap opening feeds back on the 
    $F_{1}$-fermion spectrum, violating $\omega/T$-scaling and $T$-linear resistivity.}
    \label{fig4:scaling_f}
\end{figure}

To clarify the origin of this crossover in these terms, we plot $T^{1-\eta}A_{f}(\omega,T)$ 
as a function of $\omega/T$ for $U_{fc}=1$ in Fig.~\ref{fig4:scaling_f}.  For $|\omega/T| \geq 0.05$, 
very good ``quantum critical'' scaling is visible.  For smaller $|\omega/T|$, 
however, it is clearly violated.  This violation of $\omega/T$-scaling is related to the energy scale of the $c$-fermion pseudogap which, at low energy, progressively ``feeds back'' into $A_{F_{1}}(\omega,T)$.  The
influence of the latter on strange-metal-like singularities is rather direct:
above the PG energy, IR singularities and $\omega/T$-scaling remain unaffected, and we expect $\rho(T)\simeq T$, as indeed seen in our results.  For $T$ below this PG energy scale, 
the $c$-fermion PG seems to cut-off the IR singularity in $A_{f}(\omega,T)$, 
and the influence of this PG is reflected in a $S$-shaped form of $\rho(T)$ (this is indeed a characteristic of the influence of a low-energy PG, and is seen
in many real systems, see below).  This link also suggests that clean linear-in-$T$ resistivity will obtain down to $T=0$ when the $c$-fermion PG closes.  But
in our model, this requires $U_{fc}=0$, and so we cannot reach this limit.  In a non-self-consistent version, we just have
two decoupled sectors: the strange metallic $F_{1}$ sector and the pseudogapped $c$-sector.  In this case, the total resistivity 
is {\it always} cleanly linear in $T$, since there is no dynamical feedback of the $c$-fermion PG on the $F_{1}$ self energy.
While one may be tempted to call this a realization of the strange metal, it is clearly problematic to ignore the self-consistent
feedback of the pseudogap on the singular part and vice versa.
We now point to a few real cases where such evolution as we find is (or could be) visible.

\vspace{1.0cm}

{\bf Comparison with Real Correlated Materials}

\vspace{1.0cm}

      Magic angle TBLG hosts a flat band straddling the Fermi energy, with 
gapped Dirac-like dispersive bands away from it.  There are a variety of materials that exhibit the $S=1$ Dirac cone structure, with flat and Dirac-like dispersive bands meeting in a ``triple'' crossing point at or in the proximity of $E_{F}$.  It is thus not possible to directly relate our model to these, unless the dispersive bands can be ``engineered'' to cross $E_{F}$ whilst leaving the flat band at $E_{F}$: this seems to be a tall order in practice, but is not totally inconceivable in engineered settings.  Turning to other cases of interest, a FKM-like model was derived from a full three-band Hubbard model~\cite{Varma} for cuprates in 1994.  In the context of multi-orbital systems, it is now appreciated that an orbital-selective
Mott phase (OSMP)~\cite{OSMP}, characterized by co-existing metallic and Mott insulating carriers, can widely emerge.  In the OSMP, restriction to phases with no conventional (Landau) symmetry-broken phases pre-empts descriptions in terms of conventional
Landau quasiparticles.  Such a metal is a fractionalized Fermi liquid (FFL), with itinerant fermions co-existing with local moments which, by themselves may subsequently lead to ordered states or continue to remain in spin liquid states.  In this situation, the local moment sector is associated with incoherent or Mott insulating fermions.  A hard gap or pseudogap characterizes this sector:
in underdoped cuprates, cluster-DMFT studies show that selectively Mott localized anti-nodal fermions co-exist with itinerant 
nodal fermions~\cite{Bacq-Lebreuil}.  This leads to a two-fluid model.  

The link between the strange-metal-like to a Landau FL-like evolution of $dc$ resistivity and pseudogap (PG) opening is well known 
in underdoped cuprates, and suggests that the $d$-wave PG cuts off strange metallicity in that case: the latter is recovered when the PG closes
around a so-called optimal doping, where $T_{c}$ also maximizes.  Remarkably, similar evolution of the resistivity as a function of twist angle is seen in TBLG (in this case, it is, however, unclear whether a two-fluid description or a PG can be invoked).  But {\it if} we identify the dispersionless $F_{1}$-fermion states with localized AN states and dispersive $c$-fermions with nodal (N) states, or with the two fermionic components found by Sire {\it et al.} (Varma) in cuprates, our results could be fruitfully applied.  In fact, a direct comparison between our results and those of Barisic {\it et al.}~\cite{Barisic} (see their Fig.(2) and our result for $\rho(T)$ above) shows very good qualitative accord as regards the {\it shape} of $\rho(T)$.  However, the crossover scale from the high-$T$ linear-in-$T$ to a low-$T$ pseudogapped behavior in our toy model cannot be compared with data: it is sizably higher than in UD cuprates.  Moreover, our metal results from a band-width, rather than a doping-driven Mott transition: thus, any sensible comparison should focus on pressure-driven MI transitions, as in alkali fullerides (see below) and BaFe$_{2}$S$_{3}$~\cite{bafe2s3}, and this may also apply to possible pressure-driven Mott transitions in cuprates~\cite{Baskaran}.   In these cases, we need to associate increasing doping with reduced $U_{fc}/W$ in our model.  For example, the fact that holes doped into the cuprate Mott insulator lead predominantly to appearance of nodal states (in both experiment and CDMFT), at least up to a certain doping, translates into an increase in $W$ in our model.  But the coherent band-width of nodal states (the anti-nodal states predominantly remain in selectively Mott localized and/or valence bond singlet states~\cite{Bacq-Lebreuil}), $W$, scales with the hole density, and is small in the underdoped case.  Given this, a small-to-modest $U_{fc}$ in our toy model, necessary to satisfy $U_{fc}/W\simeq O(0.5-1.8)$, is sufficient.  Moreover, with the consequent screening-induced reduction of $U_{fc}$, this translates into a $U_{fc}/W$ that reduces with increasing doping: we expect a similar trend as a function of pressure.  If we take this view, the accord we find may provide a qualitative rationalization for transport in underdoped cuprates in terms of a two-fluid or FFL picture.  Our conclusion about a pure linear-in-$T$ resistivity occuring don to $T=0$ at a pseudogap closing ``QCP'' is also consistent with this, though we cannot describe this limiting point in a sensible way in our model.  Surprisingly, similar evolution of the resistivity is also seen in TBLG~\cite{TBLG} as a function of twist angle.

  It is quite interesting that a direct bandwidth-controlled transition from a Mott insulator to a $s$-wave superconductor obtains in pressurized alkali fullerides~\cite{Ming-QiangRen}.  In $A_{3}C_{60}$ fullerides, the Jahn-Teller distortion (JTD) is also crucial and, interestingly, it's effect is to split the $t_{2g}$-orbital degeneracy~\cite{IwaharaChibotaru}.  Beyond a Hubbard $U\simeq 0.75$~eV, the JTD suddenly increases 
concomitant with the above splitting, suggesting onset of (partial) electron localization.  Once this happens, the two electrons in the lower-lying, two-fold degenerate orbital sector can undergo selective Mott localization, leaving the third electron in the higher, non-degenerate orbital in a metallic state.  In this OSMP, the former can readily give a local RVB-like correlation, inducing a low-energy gap (or pseudogap).  Though ours is a toy model of the actual, complicated situation, the above arguments suggest that it is likely that 
a situation alike the two-fluid situation we have studied can arise in fullerides.  DMFT work~\cite{Capone} indeed shows
that the ``normal'' state above $T_{c}$ can be viewed as a two-component system
with two scales: one ($T_{+}$) corresponds to a metallic component, while the other ($T_{-}$) corresponds to a correlation-induced, selectively (Mott) localized
component.  For $(U/W)>(U/W)_{c}=0.82$, we expect a pseudogapped metal with orbital-selective Mott localization to obtain.  This state seems to evolve into a ``strange''-like metal when $T_{-}=0$, exactly at $(U/W)_{c}$.  It would be interesting to see whether the $dc$ resistivity in this two-fluid metal bears 
resemblance to our result.

      Moreover, a pressure-induced Mott transition, followed by poor metallicity and possible superconductivity with a dome-like dependence on $P$ has been 
      seen in BaFe$_{2}$S$_{3}$~\cite{bafe2s3}.  Interestingly, a high $T$ linear-in-$T$ to a lower $T$ pseudogap like dependence (for $P<P_{c}$, a critical pressure less than that needed to achieve ``optimal'' $T_{c}$) obtains here as well.  $\rho_{dc}(T)\simeq \rho_{0}+AT$ seems to obtain at 
      $P_{c}$, and a superlinear $\rho_{dc}(T)\simeq T^{n}$ law with $n>1$ seems
      to obtain for $P>P_{c}$.  This is quite similar to the doping-evolution of $\rho_{dc}(T)$ in cuprates, except for the fact that pressure is the tuning parameter.  An OSMP and resulting two-fluid behavior is generic to 
      Fe arsenides.  Our toy model could also serve as a representation of the 
      $P<P_{c}$ state, and it would be interesting to see if this state hosts a low-energy pseudogap in spectral probes.

      Finally, it is interesting to wonder whether such a two-fluid scenario
      can apply to real situations involving an interplay between FB topology,
      wider band itinerance and local atomic correlations~\cite{Coleman}.
      
\vspace{1.0cm}

{\bf Spin-Charge Separation}

\vspace{0.5cm}

     An especially novel proposal of Anderson was that strange metallicity in 
cuprates bears an intimate link to spin-charge separation~\cite{Anderson}.  In 
this section, we discuss a high-dimensional realization of this exotic scenario
 within our toy model.

     The local dynamical spin susceptibility found before (this is just the 
``excitonic'' singularity in the X-ray edge problem) reads

\be
\chi_{\uparrow\downarrow}^{cF_{1}}(\omega)\simeq |\omega|^{-(2\eta-\eta^{2})}
\ee

with $\eta=(\delta/\pi)$ is the X-ray edge singularity exponent (with $\delta=$tan$^{-1}(U_{fc}/W)$ being the scattering phase shift).  At finite $T$, the dynamical spin 
susceptibility will generically show $\omega/T$-scaling: 

\be
T^{(2\eta-\eta^{2})}\chi^{cF_{1}}(\omega,T)\simeq F(\omega/T)
\ee
 
     On the other hand, at low energy, the dynamical {\it charge} fluctuation propagator in the $q\rightarrow 0$ limit is estimated to be

\be
\chi_{ch}(q,\omega)\simeq \frac{1}{(\omega/z_{F_{1}}(\omega))+iD_{F_{1}}q^{2}}
\ee

Here, (see Supplementary Information for details) 
 we use the fact that the $F_{1}$-fermion species possesses, as  
emphasized above, a 
non-trivial dynamics arising from X-ray edge physics, and that Im$\Sigma_{F_{1}}(\omega)\simeq -c_{1}-c_{2}|\omega|$
at low energy around $E_{F}(=0)$.  
This is crucial, because $z_{F_{1}}(\omega)$ is no longer a finite 
constant as in Landau FL theory, but vanishes like $-[ln\omega]^{-1}$ at the 
``Fermi surface'' (remember, though, that there is no well-defined FS in our 
toy model because of the finite Im$\Sigma_{c}(\omega=0)$).  We find that
 an appropriate Ward-Takahashi identity for the lattice model in 
high dimensions induces branch-point singular behavior in the 
charge fluctuation propagator.  Using the ``quantum hydrodynamic'' relation, we have

\be
\sigma(\omega)= Lim_{q\rightarrow 0}\frac{\omega}{q^{2}}Im\chi_{ch}(q,\omega)\simeq \frac{1}{\omega ln^{2}\omega}
\ee

     This is explicitly non-Drude-like as well.  Since the dynamical feedback of the $c$-fermion pseudogap ($c$-PG) reinstates low-energy quasicoherence in the $dc$resistivity, we expect that the above incoherent optical response will also be cut off at energies much smaller than the $c$-PG itself.

     But there is also the contribution from the $c$-fermions.  As for the usual FK model case, this contribution in our case reads

\be
\chi_{c}^{(ch)}(q,\omega)\simeq \frac{1}{\omega z_{c}^{-1}(\omega)+iD_{0,c}q^{2}}
\ee
which, notwithstanding absence of Landau quasiparticles, still exhibits a 
linear-in-$\omega$ dependence in Im$\chi_{c}^{(ch)}(\omega)$ at low energy~\cite{Freericks-RMP}.  Thus, as expected, the charge fluctuations also reflect orbital-selective Mottness: the ``itinerant'' and selectively-localized components show qualitatively distinct behavior.  In general, both will contribute.
     
     Thus, at energies above a low-energy scale associated with the influence of the $c$-PG on the strange metal-like $F_{1}$-sector, we arrive at a high-dimensional manifestation of Anderson's spin-charge 
separation, in the sense that locally, the charge and spin correlations decay 
with distinct exponents.  In the ``normal'' state, this prevents coherent 
one-fermion (Landau quasiparticle) propagation.  In principle, it also makes 
it possible to have separate instabilities to spin and charge order.

    Such singular correlations can mediate a novel, non-BCS instability.  In 
our case, as shown above, both, charge and spin fluctuations are infra-red singular: if the pair formation scale is higher than the low-energy scale we found above, this means that the SC pair correlator will also be intrinsically 
enhanced in a ``quantum critical'' sense.  Whether such SC instabilities,
 now necessarily non-$s$-wave, obtain in our case upon coupling neighboring
 local ``impurities'' is thus a very attractive issue.  In case of incipient 
 instability to a $d$-wave SC (see below) , we would need to couple four such 
``impurities'' within cluster-DMFT.  This suggests a link to cellular
 DMFT approaches~\cite{Werner}, but cementing it calls for much more work.

\vspace{1.0cm}

{\bf Superconductivity in the Toy Model}

\vspace{0.5cm}

     We now discuss the instability of the above incoherent metal to a superconductor at lower $T$.

     The strange metal-like features found here in the toy model suggest that 
any eventual instability to a superconductor should not be of the conventional 
BCS type.  First, due to the rigorous local $U(1)$ symmetry of the $S=1/2$ $H_{FK}$, due to $[n_{i,F_{1}},H]=0$ for all $i$, any 
order parameter not invariant under this symmetry must vanish by Elitzur's 
theorem.  This implies that the local component of the pairing amplitude, 
$\Delta_{ii}=<c_{i\uparrow}^{\dag}c_{i\downarrow}^{\dag}>=0=<F_{1,i,\uparrow}^{\dag}F_{1,i,\downarrow}^{\dag}>$, and so the 
SC must have gap function nodes in both $c,f$ sectors.  More crucially, there are no normal state 
long-lived electron-like quasiparticles at all, precluding any conventional 
route to the instability.  Rather, the elementary excitations are 
$(i)$ Hubbard-band pseudoparticles, $(1-n_{i,F_{1}})c_{i,\sigma}, n_{i,F_{1}}c_{i,\sigma}$, which behave neither as 
fermions, nor as bosons.  In fact, because $(1-n_{i,F_{1}\sigma'})c_{i,\sigma}=(F_{1,i,-\sigma'}S_{i,\sigma\sigma'}^{-}+F_{1,i,\sigma'}S_{i}^{+})$ with $S_{i,\sigma\sigma'}^{-}=F_{1,i,-\sigma'}^{\dag}c_{i\sigma}$, etc, the elementary excitations
 are composites of an (unprojected) fermion and a local ``excitonic'' charge or spin fluctuation (the
 latter are hard-core bosons), and $(ii)$ a conformally invariant infra-red 
singular continuum of multifermion character, given by 
Im$G_{ii,F_{1}}(\omega)\simeq \omega^{-(1-\eta)}$.  Such a composite 
operator has {\it no} overlap with any Landau quasiparticle (whence our finding of $z_{F_{1}}=0$ at $E_{F}$ above).  Finally, given incoherence of the $c,F_{1}$-fermion states, strong quantum phase fluctuations must
be involved (just because of the number-phase uncertainty principle) in any eventual spin-singlet pair condensation.

       In a two-fluid picture of co-existing metallic nodal (N) and insulating 
 anti-nodal (AN) states, the crucial observation is that coherent one-electron
 mixing between these two sectors is quenched in the underdoped cuprates because of (momentum-selective) Mott localization of the AN states.  In fact, this is 
the starting point for our toy model with $c,F_{1}$ fermions associated with $N,AN$ states.

\vspace{1.0cm}

{\bf Two-Particle Residual Interaction and Pair Glue}

\vspace{0.5cm}

     To proceed, we need to generate a relevant two-fermion {\it effective} interaction.  It is important to
 re-emphasize that the one-fermion hybridization remains incoherent in the ``normal'' state above.  However, notwithstanding this, there is nothing 
to prevent onset of coherence in two-fermion hopping processes.  In fact, 
this state of affairs is quite well known in coupled $d=1$ Luttinger liquids.
  There, coherent tunnelling of one-electron quasiparticles is blocked by 
{\it precisely} the orthogonality catastrophe (when expressed in 
bosonic language, the fundamental excitations are in fact separate spin- and
 charge collective modes, the tomonagons~\cite{Anderson}.  Simply put, in $d=1$, 
the spinon and holon cannot recohere to hop as a
coherent electron-like quasiparticle because of spin-charge separation.  In our
 high-$d$ view, we do find a spin-charge separation, again simply because the 
quasiparticle weight, $z_{F_{1}}(\omega)\simeq \omega^{\eta}$ 
or -$(ln\omega)^{-1}$, 
vanishes at $E_{F}(=0)$ as a consequence of the Nozieres de-Dominicis effect in 
DMFT.  The {\it only} way for such an incoherent state to relieve it's finite 
residual entropy $O($ln$2)$ per site (in DMFT, corresponding to a partially 
unquenched, critically fluctuating local moment) is to generate 
direct instabilities to some kind of ordered state.

\vspace{0.5cm}

     To ``derive'' the residual interaction in the ``strange-like'' metal 
we find, we extend the above idea to our local limit.  We draw upon an analogy 
with what happens in coupled $d=1$ Luttinger liquids, where Anderson argues 
that in the limit where {\it coherent} one-electron hopping scales 
to irrelevance~\cite{Kusmartsev}, two-particle hopping processes in the particle-hole (ph) and particle-particle (pp) channels become more relevant.  If we consider our local limit, we replace the two chyains by two, local ``impurities''.  Now, coherent ({\it inter-site}) one electron mixing scales 
to irrelevance, thanks to vanishing Landau quasiparticle residue.  To leading 
order in $1/d$, the self-same intersite one-electron hybridization, however, generates an 
{\it effective} two-particle {\it residual interaction} by a second-order-in-$V_{fc}$ hopping process,

\be
H_{res}\simeq -\frac{1}{U}\sum_{<i,j>,\sigma,\sigma'}V_{ij}^{2}(c_{i\sigma}^{\dag}F_{1,j\sigma}+h.c)(c_{j\sigma'}^{\dag}F_{1,i,\sigma'}+h.c)
\ee

Notice that this is a two-fermion hopping process akin to that which produces the Anderson super-exchange in one-band Hubbard
model(s).  Here, however, it is a second-order-in-hybridization process, and corresponds to an anomalous multiparticle mixing between the ``dark''- and $c$-fermions (the dark fermion, $F_{1,\sigma}$, is a consequence of selective Mottness and has no single-fermion interpretation).  It becomes relevant only when coherent one-electron 
hybridiation is made irrelevant.  In our case, it is impossible to coherently mix the $c$ and $F_{1}$ fermions since, as we found above, they are
not long-lived enough to permit a coherent one-fermion transfer to occur.  

It is easy to see that $H_{res}$ contains terms like 
$c_{i\sigma}^{\dag}c_{j\sigma}F_{1,i,\sigma'}^{\dag}F_{1,j,\sigma'}$ and $c_{i\sigma}^{\dag}c_{j,-\sigma}^{\dag}F_{1,j,-\sigma'}F_{1,i,\sigma'}$.  
In DMFT, both can be decoupled in a static-HF-Bogoliubov sense, directly 
yielding two instabilities.
The first term yields a ph-order parameter, $\Delta_{ph}=\epsilon_{ij}<c_{i\sigma}^{\dag}c_{j\sigma}>, \epsilon_{ij}<F_{1,i,\sigma'}^{\dag}F_{1,j,\sigma'}>$ with $\epsilon_{ij}=+1$ for $j=i\pm x$ and $-1$ for $j=i\pm y$, leading to $d$-form 
factor density wave ``excitonic'' instabilities on a $d=2$ square lattice.
  The second gives $\Delta_{pp}=\epsilon_{ij}<c_{i\sigma}^{\dag}c_{j,-\sigma}^{\dag}>, 
  \epsilon_{ij}<F_{1,j,-\sigma'}F_{1,i,\sigma'}>$, which is precisely (in general, a nodal) a $d$-wave pair order parameter in $d=2$, because the local components of the gap functions rigorously vanish in the OSMP, see above.  Since 
these order parameters emerge from the same $H_{res}$, they naturally represent
 {\it competing} orders.  This implies that a $d$-wave ph (excitonic) 
condensate, indicative of $d$-form factor density wave, is a leading 
competitor of $d$-wave SC.  There is substantial evidence that an intracell 
charge nematic with $d$-wave bond-modulated density-wave order exists in the 
underdoped (pseudogap) regime of hole-doped cuprates~\cite{FujitaPNAS}.

     There is, in general, also a spin-singlet or triplet excitonic condensate, characterized by a finite $<c_{i,\sigma}^{\dag}F_{1,i,\sigma'}>$, as well as 
     a spin triplet pair condensate, $<c_{i,\sigma}^{\dag}F_{j,\sigma'}^{\dag}>$
     that may arise from $H_{res}$ under appropriate conditions.  Depending on the peculiarities of the $k$-space form factor of $V_{fc}(k)$, this may lead to $p$- (for $V_{fc}(k)\simeq$ sin$(k_{x})$) or $d$-wave pair symmetry
     (for $V_{fc}(k)\simeq ($cos$k_{x}-$cos$k_{y})$).  We will not consider these cases here, though they are rather interesting in their own right~\cite{Coleman}.
     
\vspace{1.0cm}

{\bf Continuum Pair Glue and Instabilities of the Strange Metal}

\vspace{1.0cm}

     We start by noticing that the infra-red singular local spin fluctuation 
(or excitonic) correlator found earlier is a natural candidate for the 
``pair glue'' that instigates a direct transition to a $d$-wave SC from the ``normal'' state described above.  In fact, we have

\be
Im\chi_{cF_{1}}^{\sigma\sigma'}(\omega)\simeq (\omega_{c}/\omega)^{\gamma_{\sigma\sigma'}}
\ee
with $\gamma_{\sigma\sigma'}=(2\delta_{\sigma\sigma'}/\pi)-(\delta_{\sigma\sigma'}/\pi)^{2}$ is the scattering phase shift arising
from the local X-ray edge physics, and $\omega_{c}$ is a high-energy cut-off.
  Assuming spin fluctuations to be more relevant, the pair glue that enters the strong coupling Eliashberg equations is then
$\alpha^{2}F(\omega)=J^{2}$Im$\chi_{cF_{1}}^{\uparrow\downarrow}(\omega)$.

If we use the optical conductivity, the pair glue is

\be
\alpha^{2}F(\omega) \simeq C\frac{\omega_{p}^{2}}{4\pi}\frac{\partial^{2}}{\partial\omega^{2}}(\omega Re\frac{1}{\sigma(0,\omega)})
\ee

\vspace{1.0cm}

          We can use either choice as an input into the Eliashberg equations.
This has been done in recent times, and we refer to extant results.
  According to Miao et al.~\cite{1805.10280v2}, a particularly notable 
consequence of such an anomalous glue is that the ratio $y=2\Delta_{0}/k_{B}T_{c}$ is enhanced over it's BCS value of 3.52.  Interestingly, for $\eta\simeq 0.3-0.4$ in Im$\chi_{cF_{1}}^{\sigma\sigma'}(\omega)$, we find $y\simeq 3.7-4.0$.  

If we use the optical conductivity, the pair glue is
 $\alpha^{2}F(\omega)\simeq C\omega_{p}^{2}/2\pi$.  
  This is independent of $\omega$: interestingly, such a possibility has been phenomenologically considered by Norman {\it et al.}~\cite{Norman-Chubukov}.  
Since the $d$-wave 
BCS value of $y$ is $4.2$, the actual value of $y$ in our case is 
$y=(3.53-4.0)(4.2)/(3.53)=4.2-4.5$, depending upon whether we use the optical conductivity or spin fluctuations to extract the pair-glue.  Although this is way off from the observed values 
of $y\simeq O(7-8)$, this is quite a remarkable result if we interpret it 
consistently.

     The point is that the Eliashberg formalism is still a mean-field theory,
notwithstanding it's sophistication.  It completely ignores quantum phase
fluctuations which must depress the true $T_{c}$ from it's mean-field value,
especially in $d=2$, and when SC arises from an {\it incoherent} normal state.  So the above value
of $y\simeq 4.2-4.5$ must be interpreted carefully.  Specifically, the $T_{c}$ must now be identified as
$T_{c}^{mf}=T_{p}$, a scale at which pairs first {\it form}, and not as that at
which global pair coherence obtains.  If we do this, excellent accord with
extant electronic Raman scattering (ERS) data~\cite{Sacuto}, resistivity, ARPES and tunnelling (STS) data~\cite{Fischer} is directly seen: indeed, $2\Delta_{pg}/k_{B}T_{p}=4.3$ is deduced from all these 
probes.  This is in excellent accord with our estimate.  This means that using
the strange metal responses,
we obtain the dominant feature associated with the instability of the strange
metal to a pseudogapped, {\it nodal} metal.  In our formulation, 
this ``instability'' is to a preformed $d$-wave paired state.  It is also 
naturally consistent with observation of precursor diamagnetism in the range 
$T_{c}<T<T_{p}$~\cite{Ong}.  It might be interesting to investigate occurence of related features in alkali-fullerides and BaFe$_{2}$S$_{3}$ under pressure, and it's implications for 
(hitherto uninvestigated) ``hidden'', competing electronic order in the pseudogap
in those cases.

      If this holds, we must conclude that the actual SC transition involves a lower scale, $T_{c}<T_{p}$, where the preformed pairs as above acquire macroscopic quantum phase coherence.  Computing quantum phase fluctuation effects using the propagators $G_{c}(\omega), G_{F_{1}}(\omega)$ is obviously of interest, and is left for the future.

\vspace{1.0cm}

{\bf Discussion}

\vspace{1.0cm}

     We have considered the issues of strange metallicity and it's instability to a $d$-wave SC using a toy model.  The important question is ``Can similar physics obtain in the doped Mott insulator in two dimensions?  {\it If} the one band Hubbard model in $d=2$ is taken to be a minimal model for cuprates, how may one imagine occurence of strange metal singularities 
that are necessary?  A way out would be to appeal to cluster extensions 
of DMFT: after all, important aspects of cuprates like ${\bf k}$-space
 differentiation of electronic states are beyond reach of DMFT.  This is indeed our motivation for the toy model for underdoped cuprates: upon associating $c=c_{k_{N}}$ and $F_{1,k}=c_{k_{AN}}$, this mimics the two-fluid situation of (Mott) localized AN states co-existing with metallic N states, but
only above $T=T_{p}$.  Below $T_{p}$, a $d$-wave pseudogap, inaccessible in
{\it any} local approximation, opens up.  While we cannot describe the physics below $T_{p}$, it is quite notable that we {\it can} describe the leading mean-field instability of the ``high-$T$'' incoherent metal to such a ``preformed paired'' state.  It is also interesting to see that the crossover from high-$T$ linear-in-$T$ to a pseudogapped low-$T$ form of $\rho_{dc}(T)$ {\it is}
recovered in our model, due to a Mottness-induced $c$-fermion PG rather than a 
preformed $d$-wave gap.  This is also the reason why the incoherent-to-pseudogapped crossover in our $\rho_{dc}(T)$ occurs at a $T$ scale higher than that seen in UD cuprates: in reality, the PG arises from $d$-wave preformed-pair
correlations at a scale $O(J=4t^{2}/U$) in cuprates~\cite{Bacq-Lebreuil}, while our PG is related to pure Mottness.  Nonetheless, it is remarkable that the details of the crossover as well as T-linearity of $\rho_{dc}(T)$ appearing as a PG-closing QCP do resemble what is seen in UD cuprates.  

    The resolution of this aspect may be already present in extant 
cluster-DMFT studies.  In particular, Hoshino {\it et al.}~\cite{Werner} and Bacq-Lebreuil~\cite{Bacq-Lebreuil} have 
mapped the one-band Hubbard model on an embedded four-site cluster to a four-
``orbital'' Kanamori-Hubbard (KH) model.  Such a cluster-to-orbital mapping 
relates momentum selective Mott physics in cluster DMFT to the OSMP in the KH 
model, opening the door to such selective-Mott physics in cuprates within a 
more realistic Hubbard  model in $d=2$.  On an embedded 4-site cluster in the 
OSMP or spin-freezing regime, we expect that anti-nodal states (corresponding 
to cluster momenta $(\pm\pi,0), (0,\pm\pi)$) will be selectively localized, 
while nodal states with cluster momenta $(\pm\pi/2,\pm\pi/2)$ will remain 
metallic.  Once this obtains, we then expect such X-ray edge physics to occur at an {\it intracluster} level.  This is indeed what is seen in these 
(cluster)-DMFT simulations, where the doping-dependent exponent of the 
cluster local anti-nodal Green function~\cite{Werner} displays 
infra-red branch-point singular behavior.  Moreover, the $AN$ pseudogap is now governed by a scale $O(J\simeq 4t^{2}/U<<U)$, and this may help resolve the problem of the $T$-scale of the resistivity crossover mentioned above.

  Our analysis captures the dominantly local physics at the heart 
of the strange metal in a toy model sense.  However, the localization-delocalization 
transition of the anti-nodal states, corresponding to the topological 
transition from a ``small'' to ``large'' Fermi surface around optimal 
doping, needs 
further extension of the present approach.  We need $V_{fc}(k)$ to be RE-relevant in the infra-red to trigger a ``transition'' from the strange-metal-like state obtained above to a Landau Fermi liquid metal with a large Fermi surface that counts bothy, $c$ and $F_{1}$ fermions.  For cuprates, incorporating these aspects needs 
(cluster/cellular) DMFT studies, involving careful extension of the present 
work for the ``four-orbital'' KH model in it's momentum-selective Mott phase.
  But this may well preclude analytic insight.  
We leave this aspect for the future.

\vspace{1.0cm}

{\bf Conclusion}

\vspace{0.5cm}

     In conclusion, we have investigated the strange metal-like anomalies 
emerging in a toy model for an OSMP.  We have devised a way to go beyond the famed alloy analogy approximation (AAA) for the $S=1/2$ FKM, and discussed it's possible implications for transport in pseudogapped metals {\it without} any connection of proximity to conventional symmetry breaking.  
  We also show how the loss of Landau quasiparticles due to a 
many-body X-ray edge effect manifests itself in a high-dimensional 
spin-charge separation: both, dynamic spin and charge correlations decay 
anomalously slowly, but with different fractional exponents.  Such anomalously
 singular spin and charge continuua can be attractive candidates as 
unconventional, intrinsically multiparticle pair glues for $d$-wave 
superconductivity.

\vspace{2.4cm}

{\bf Supplementary Information}
 
\vspace{1.3cm}

{\bf Infra-red Singular behavior of $G_{F_{1}}(\omega)$ in the FKM}

\vspace{0.5cm}

     In this section, we derive the explicit form of the infra-red singular
(branch-point analytic structure) behavior of $G_{F_{1}}(\omega)$ used in
the main text.  We employ the equation-of-motion technique.  The EOM for the $F_{1}$-fermion is

\be
\omega G_{F_{1}}(\omega)=1+U_{fc}\langle n_{i,c}F_{1,i};F_{1,i}^{\dag}\rangle
\ee
where there is no hybridization ``bath'' function term, because the $F_{1}$-fermions do not hop.  One may include the Hartree 
contribution in the above, simply by subtracting the term $U_{fc}n_{c}$ 
with $n_{c}=(1/N)\sum_{i}\langle n_{i,c}\rangle$ ($N$ is the number of sites) on both sides, with the result

\be
(\omega-U_{fc}n_{c,\uparrow})G_{F_{1}}(\omega)=1+U_{fc}\langle (n_{i,c}-n_{c})F_{1,i};F_{1,i}^{\dag}\rangle
\ee

     In Zubarev's~\cite{zubarev} method, one writes down a new EOM for the GF appearing on the RHS in the above equation.  Of course, this leads to an infinite chain of EOMs
that generate progressively higher-order GFs, and a suitable decoupling is then 
necessary to close the chain of these EOMs to obtain $G(\omega)$.  We take a 
different route, based on differentiating w.r.t the second time variable ($t'$)
in the two-time GF, $G_{F_{1}}(t-t')=-i\theta(t-t')\langle F_{1}(t);F_{1}^{\dag}(t')\rangle$~\cite{IGF}.  This leads to

\be
\omega \langle (n_{i,c}-n_{c})F_{1,i};F_{1,i}^{\dag}\rangle = U_{fc}\langle (n_{i,c}-n_{c})F_{1,i};(n_{i,c}-n_{c})F_{1,i}^{\dag}\rangle
\ee

which can be rewritten as 

\be
(\omega-U_{fc}n_{c})\langle (n_{i,c}-n_{c})F_{1,i};F_{1,i}^{\dag}\rangle = U_{fc}\langle (n_{i,c}-n_{c})F_{1,i};(n_{i,c}-n_{c})F_{1,i}^{\dag}\rangle
\ee

     Calling $\langle (n_{i,c}-n_{c})F_{1,i};(n_{i,c}-n_{c})F_{1,i}^{\dag}\rangle = \Gamma_{ii}^{cF_{1}}(\omega)$, we get

\be
G_{F_{1}}(\omega)=\frac{1}{\omega -U_{fc}n_{c}-\frac{U_{fc}^{2}\Gamma^{cF_{1}}(\omega)}{1+U_{fc}^{2}\Gamma^{cF_{1}}(\omega)G_{0,c}(\omega)}}
\ee
     where $G_{0,c}^{-1}(\omega)=(\omega -U_{fc}n_{c})$, and we drop the site index.  The $F_{1}$-fermion self energy is

\be
\Sigma_{F_{1}}(\omega)=U_{fc}n_{c}+\frac{U_{fc}^{2}\Gamma^{cF_{1}}(\omega)}{1+U_{fc}^{2}\Gamma^{cF_{1}}(\omega)G_{0,c}(\omega)}
\ee

     Eq.(50) can be trivially rewritten as

\be
G_{F_{1}}(\omega) = G_{0,c}(\omega) + G_{0,c}(\omega)U^{2}\Gamma^{cF_{1}}(\omega)G_{0,c}(\omega)
\ee
     enabling us to identify $T(\omega)=U_{fc}^{2}\Gamma^{cF_{1}}(\omega)$ as a
``scattering $T$-matrix'': it is a three-fermion correlator.  We notice that the self-energy is composed of contributions from this scattering $T$-matrix 
to all orders in $U_{fc}$.  
Focussing on $\Gamma^{cF_{1}}(\omega)$, we can decouple this local {\it three-fermion} propagator as follows:

\be
U_{fc}^{2}\Gamma^{cF_{1}}(\omega_{n})=(U_{fc}/\beta)^{2}\sum_{\omega_{1},\omega_{2}}G_{c}^{(0)}(\omega_{1})G_{c^{(0)}}(\omega_{2})G_{F_{1}}^{(0)}(\omega_{n}-\omega_{1}-\omega_{2})
\ee
Using the spectral representation, $G_{c}^{(0)}(\omega_{n})=\int\frac{\rho_{0,c}(\epsilon)}{i\omega_{n}-\epsilon}$ and doing
the Matsubara sum yields

\be
U_{fc}^{2}\Gamma^{cF_{1}}(\omega_{n})=(U_{fc}/2)^{2}\int_{-\infty}^{\infty}d\epsilon_{1} d\epsilon_{2}\frac{\rho_{0,c}(\epsilon_{1})\rho_{0,c}(\epsilon_{2})}{i\omega_{n}-\epsilon_{1}-\epsilon_{2}}tanh(\beta\epsilon_{1}/2)[tanh(\beta\epsilon_{2}/2)+coth(\beta\epsilon_{1}/2)]
\ee
Analytic continuation, $i\omega_{n}\rightarrow \omega$, and approximating $\rho_{0,c}(\epsilon)\simeq \rho(E_{F})=\rho(0)$ (this is thus valid at low energies), we obtain

\be
U_{fc}^{2}Im\Gamma^{cF_{1}}(\omega,T)=-\frac{\pi}{2}(\frac{U_{fc}}{\pi W})^{2}\omega coth(\beta\omega/2)
\ee
yielding it's real component as

\be
U_{fc}^{2}Re\Gamma^{cF_{1}}(\omega,T)=(\frac{U_{fc}}{\pi W})^{2}\omega ln(\frac{max{[\omega,T]}}{W})
\ee

yielding a Landau QP residue, $z(\omega)\simeq -$ [ln$\omega]^{-1}=0$; {\it i.e}, a marginal FL form.  The $F_{1}$-fermion self-energy above thus corresponds to a sum of such (logarithmic) terms to infinite order, and this presages the infra-red power-law form we wish to get.

\vspace{0.5cm}

     Non-trivial changes occur as soon as the $F_{1}$-fermion acquires a non-zero hopping (finite mass) via finite $V_{fc}$.
     The lack of a local degeneracy in the impurity problem when $V_{fc} \neq 0$ immediately cuts-off
the infra-red divergent excitonic susceptibility.  In terms of diagrams, this occurs because the recoil of the
$F_{1}$-fermion reinstates the standard phase space argument when one
 considers scattering of $c$ and $F_{1}$ fermions at $E_{F}$.  This must lead directly to re-appearance of severely renormalized (depending upon $V_{fc}/U_{fc}$ and the band-filling) Landau quasiparticles~\cite{Toulouse1971}.  This is indeed what happens in the Hubbard model, or in the multiband Kanamori-Hubbard models or the EPAM in their non-OSM phases.

      This demonstrates the one-to-one link between selective-localization and 
emergent infra-red singular spectral responses.  Though this is harder to show analytically in finite $U_{ff}$ multi-band Hubbard models or in cluster-DMFT studies for the 
one-band Hubbard model in $d=2$, this one-to-one link should continue 
to hold in orbital- or momentum-selective Mott phases in these models, because 
the above arguments only require {\it co-existent} itinerant and localized states at the Fermi surface in a (selective) metallic phase without any conventional symmetry-breaking. 
     
\vspace{1.0cm}

{\bf Dynamical Charge Susceptibility}

     The argument goes as follows: The self-consistency condition of DMFT leads to the equivalence of the Ward identity for the lattice model with
  that for the corresponding ``impurity'' model as ($a=c,F_{1}$)~\cite{GKKR1996}

\be
\Sigma_{a}(\nu+\omega) - \Sigma_{a}(\nu) = T\sum_{\nu'}\gamma^{a}_{\nu\nu'\omega}[g_{a}(\nu'+\omega)-g_{a}(\nu')]
\ee
where $\gamma_{a}$ is the two-particle self-energy of the impurity model.  Using DMFT selfconsistency, we have $g_{a}(\nu)=(1/N)\sum_{k}G_{a}({\bf k},\nu)$, where $G_{a}({\bf k},\nu)$ are the one-fermion propagators we found in the main text.  Then,

\be
\Sigma_{a}(\nu+\omega)-\Sigma_{a}(\nu) = \frac{T}{N}\sum_{k',\nu'}\gamma^{a}_{\nu\nu'\omega}G_{a,k',\nu'}G_{a,k',\nu'+\omega}[i\omega -(\Sigma_{a}(\nu'+\omega)-\Sigma_{a}(\nu'))]
\ee

and thus,

\be
\frac{\Sigma_{a}(\nu+\omega)-\Sigma_{a}(\nu)}{i\omega} =\frac{T}{N}\sum_{k',\nu'}\gamma^{a}_{\nu\nu'\omega}G_{a,k',\nu'}G_{a,k',\nu'+\omega}[1-\frac{\Sigma_{a}(\nu'+\omega)-\Sigma_{a}(\nu')}{i\omega}]
\ee

  On the other hand, for any arbitrary ${\bf q}$, the Bethe-Salpeter eqn for the
  two-particle vertex reads

\be
F_{\nu\nu'}^{a}({\bf q},\omega)=\gamma^{a}_{\nu\nu'\omega} + \frac{T}{N}\sum_{k'',\nu''}\gamma^{a}_{\nu\nu''\omega}G_{a,k'',\nu''}G_{a,k''+q'',\nu''+\omega}F_{\nu''\nu'}^{a}({\bf q},\omega)
\ee

 We multiply this eqn by $G_{a,k'}G_{a,k'+q}$ and sum the result over ${\bf k}',\nu'$ to obtain

\be
\frac{T}{N}\sum_{k',\nu'}G_{a,k',\nu'}G_{a,k'+q,\nu'+\omega}F_{\nu\nu'}^{a}({\bf q},\omega)=\frac{T}{N}\sum_{k',\nu'}\gamma^{a}_{\nu\nu'\omega}G_{a,k'\nu'}G_{a,k',\nu'+\omega}[1+\frac{T}{N}\sum_{k'',\nu''}G_{a,k'',\nu''}G_{a,k''+q'',\nu''+\omega}F_{\nu''\nu'}^{a}({\bf q},\omega)]
\ee

Comparing equations (37) and (39), we see that they actually represent the {\it same} eqn.  Hence,

\be
-\frac{\Sigma_{a}(\nu+\omega)-\Sigma_{a}(\nu)}{i\omega} = \frac{T}{N}\sum_{k'',\nu''}G_{a,k'',\nu''}G_{a,k''+q'',\nu''+\omega}F_{\nu'',\nu'}^{a}({\bf q},\omega)
\ee

     Hence, the three-leg vertex actually varies as the inverse of the quasiparticle residue:

\be
\Lambda_{a}({\bf q},\omega) = 1 - \frac{\Sigma_{a}(\nu+\omega)-\Sigma_{a}(\nu)}{i\omega} \rightarrow z_{a}^{-1}(\omega)
\ee

We now use our DMFT result, where we found that Im$\Sigma_{F_{1}}(\omega)\simeq |\omega|^{1-\eta}$ or -$|\omega|$ and Im$\Sigma_{c}(\omega)$ having the ``wrong'' sign at low energy.  This gives $z_{F_{1}}(\omega)\simeq \omega^{\eta}$ or -$(ln\omega)^{-1}$, leading to an infra-red divergence of the three-leg vertex,
$\Lambda(\omega)\simeq \omega^{-\eta}$ or $\simeq -$ln$\omega$.  Thus, the singular fermion self-energies at the MIT within DMFT directly lead to
infra-red singular $F_{1}$-fermion vertex, and the latter leads to drastic modification of the low-energy charge fluctuation spectrum.  Explicitly~\cite{Vignale-Hanke},

\be
\chi_{ch}^{F_{1}}(q,\omega) \simeq \frac{1}{\omega.z_{F_{1}}^{-1}(\omega)+iD_{0,F_{1}}q^{2}}
\ee

If $D_{0,F_{1}}\simeq 0$, this leads to a momentum-independent charge response.

But as in the usual FK case~\cite{Freericks-RMP}, the $c$-fermion charge susceptibility
still varies linearly with $\omega$ at low energy, notwithstanding absence of Landau quasiparticles.  Explicitly,

\be
\chi_{ch}^{(c)}(q,\omega)=\frac{1}{\omega.z_{c}^{-1}(\omega)+iD_{0,c}q^{2}}
\ee

Thus, as one would expect in a two-fluid picture, we find qualitatively distinct charge fluctuation responses in the itinerant and selectively Mott-localized sectors in the OSMP.  Notwithstanding it's incoherent propagator, the $c$-sector exhibits the linear-in-$\omega$ form of the {\it local} charge-fluctuation spectrum, while that of the $F_{1}$-sector is a highly anomalous continuum.

Thus, the pole structure of the $F_{1}$-fermion diffusion propagator in a correlated Landau Fermi liquid with a finite Landau quasiparticle residue,
$0 < z <1$, is supplanted by a branch-point singularity in our case. 
 In a Landau Fermi liquid (LFL), $z(\omega)=z$, a constant in the infra-red, 
leading to regular diffusion modes.  Here, it is the anomalously vanishing LFL 
quasiparticle residue that leads to anomalous diffusive behavior, characteristic of ``anomalous quantum hydrodynamics''.  This leads
directly to the anomalously slow power-law fall off in optical conductivity as
detailed in the main text.

\bibliographystyle{apsrev4-2}
\bibliography{strange}

\end{document}